\documentclass[10pt,journal,compsoc]{IEEEtran}
\pdfoutput=1
\usepackage{cite}
\usepackage{amsmath,amssymb,amsfonts}
\usepackage{graphicx}
\usepackage{textcomp}
\usepackage{xcolor}

\usepackage{graphicx}
\usepackage{balance}  % for  \balance command ON LAST PAGE  (only there!)
\usepackage{subcaption}

\usepackage{booktabs} % For formal tables
\usepackage[ruled,lined,linesnumbered,noend]{algorithm2e}
\usepackage{color}
\usepackage[hyphens]{url}
\usepackage{array}
\usepackage{amsfonts}
\usepackage{amsthm}
\usepackage[symbol]{footmisc}

\SetAlFnt{\small}
\SetAlCapFnt{\small}
\SetAlCapNameFnt{\small}
\SetAlCapHSkip{0pt}
\IncMargin{-\parindent}

\newtheorem{definition}{Definition}
\newtheorem{example}{Example}
\newtheorem{theorem}{Theorem}

\newcommand{\predicate}[2]{{\sf #1 = #2}}

\newcommand{\slicefinder}{{\sf Slice Finder}}

\newcommand{\squishlist}{ 
   \begin{list}{$\bullet$}
    { \setlength{\itemsep}{0pt}      \setlength{\parsep}{3pt} 
      \setlength{\topsep}{3pt}       \setlength{\partopsep}{0pt}
      \setlength{\leftmargin}{1.5em} \setlength{\labelwidth}{1em}
      \setlength{\labelsep}{0.5em} } }

\newcommand{\squishend}{
    \end{list}  } 

\begin{document}

%\numberofauthors{2}

\title{Automated Data Slicing for Model Validation: \\A Big data - AI Integration Approach}

\author{Yeounoh~Chung,
        Tim~Kraska,
        Neoklis~Polyzotis,
        Ki~Hyun~Tae,
        and~Steven~Euijong~Whang,~\IEEEmembership{Member,~IEEE}% <-this % stops a space
\IEEEcompsocitemizethanks{
\IEEEcompsocthanksitem Y. Chung is with Apple Inc. Work done at Google Research.\protect\\
E-mail: yeounoh\_chung@brown.edu
\IEEEcompsocthanksitem T. Kraska is with MIT. E-mail: kraska@mit.edu
\IEEEcompsocthanksitem N. Polyzotis is with Google Research. 
E-mail: npolyzotis@google.com
\IEEEcompsocthanksitem K. Tae is with KAIST. E-mail: kihyun.tae@kaist.ac.kr
\IEEEcompsocthanksitem S. E. Whang is with KAIST. Work done at Google Research and KAIST.\protect\\E-mail: swhang@kaist.ac.kr
\IEEEcompsocthanksitem Corresponding author: S. E. Whang
}% <-this % stops a space
}

% \author{
% \IEEEauthorblockN{Yeounoh Chung\IEEEauthorrefmark{1}}
% \IEEEauthorblockA{Brown University \\
% yeounoh\_chung@brown.edu}
% \and
% \IEEEauthorblockN{Tim Kraska}
% \IEEEauthorblockA{MIT CSAIL\\
% kraska@mit.edu}
% \and
% \IEEEauthorblockN{Neoklis Polyzotis}
% \IEEEauthorblockA{Google Research \\
% npolyzotis@google.com}
% \and
% \IEEEauthorblockN{Steven Euijong Whang\IEEEauthorrefmark{1}}
% \IEEEauthorblockA{KAIST \\
% swhang@kaist.ac.kr}
% }

\IEEEtitleabstractindextext{
\begin{abstract}
As machine learning systems become democratized, it becomes increasingly important to help users easily debug their models. However, current data tools are still primitive when it comes to helping users trace model performance problems all the way to the data.
We focus on the particular problem of slicing data to identify subsets of the validation data where the model performs poorly.
This is an important problem in model validation because the overall model performance can fail to reflect that of the smaller subsets, and slicing allows users to analyze the model performance on a more granular-level.
Unlike general techniques (e.g., clustering) that can find arbitrary slices, 
our goal is to find interpretable slices (which are easier to take action compared to arbitrary subsets) that are problematic and large.
We propose \slicefinder{}, which is an interactive framework for identifying such slices using statistical techniques. Applications include diagnosing model fairness and fraud detection, where identifying slices that are interpretable to humans is crucial. This research is part of a larger trend of Big data and Artificial Intelligence (AI) integration and opens many opportunities for new research.
\end{abstract}

}

\maketitle

\IEEEraisesectionheading{\section{Introduction}\label{sec:introduction}}
%\tim{Alternative intro sketch}

\IEEEPARstart{M}{achine} learning systems~\cite{baylor2017tfx} are becoming more prevalent thanks to a vast number of success stories.  
 However, the data tools for interpreting and debugging models have not caught up yet, and many important challenges exist to improve our model understanding after training~\cite{doshi2017towards}. 
One such key problem is to understand if a model performs poorly on certain parts of the data, hereafter referred to as a {\em slice}.

\begin{example}\label{example_intro} Consider a Random Forest classifier that predicts whether a person's income is above or below \$50,000 (UCI Census data \cite{Lichman:2013}). Looking at Table~\ref{tbl:census}, the overall metrics may be considered acceptable, since the overall log loss (a widely-used loss metric for binary classification problem) is low for all the data (see the ``All'' row). However, the individual slices tell a different story. When slicing data by gender, the model is more accurate for {\sf Female} than {\sf Male} (the \emph{effect size} defined in Section~\ref{sec:problem} captures this relation by measuring the normalized loss metric difference between the {\sf Male} slice and its counterpart, the {\sf Female} slice).
%for being positive for {\sf Male} slice, but negative for {\sf Female} slice). 
The {\sf Prof-specialty} slice is interesting because the average loss metric is on par with {\sf Male}, but the effect size is much smaller. A small effect size means that the loss metric on {\sf Prof-specialty} is similar to the loss metric on other demographics (defined as {\em counterparts} in Section~\ref{sec:problem}). 
Hence, if the log loss of a slice and that of the counterpart are not acceptable, then it is likely that the model is bad overall, not just on a particular subset.
Lastly, we see that people with higher education degrees ({\sf Bachelors} $<$ {\sf Masters} $<$ {\sf Doctorate}) suffer from worse model performance, and their losses are higher than their counterparts and thus have higher error concentration. Thus, slices with high effect size are important for model validation, to make sure that the model does not underperform on certain parts of the data.
\end{example}

\begin{table}[t]
\centering
\begin{tabular}{| c | c | c | c |}
\hline
{\bf Slice} & {\bf Log Loss} & {\bf Size} & {\bf Effect Size} \\
\hline \hline
%All & 0.47 & 30k & n/a \\
All & 0.35 & 30k & n/a \\
\hline
\predicate{Sex}{Male} & 0.41 & 20k & 0.28 \\
\hline
\predicate{Sex}{Female} & 0.22 & 10k & -0.29 \\
\hline
\predicate{Occupation}{Prof-specialty} & 0.45 & 4k & 0.18 \\

%\predicate{Workclass}{Local-gov} & 0.43 & 1.7k & 0.19 \\
%\predicate{Race}{White }& & & \\
\hline
\predicate{Education}{HS-grad} & 0.33 & 9.8k & -0.05 \\
\hline
\predicate{Education}{Bachelors} & 0.44 & 5k & 0.17 \\
\hline
\predicate{Education}{Masters} & 0.49 & 1.6k & 0.23\\
\hline
\predicate{Education}{Doctorate} & 0.56 & 0.4k  & 0.33 \\
\hline
\end{tabular}
\caption{UCI Census data slices for Example~\ref{example_intro}}
\vspace{-0.2cm}
\label{tbl:census}
\end{table}

The problem is that the overall model performance can fail to reflect that of smaller data slices. Thus, it is important that the performance of a model is analyzed on a more granular level.
While a well-known problem~\cite{mcmahan2013ad}, current techniques to determine underperforming slices largely rely on domain experts to define important sub-populations (or at least specify a feature dimension to slice by)~\cite{kahng2016visual,tfma}. Unfortunately, machine learning practitioners do not necessary have the domain expertise to know all important underperforming slices in advance, even after spending a significant amount of time exploring the data.
An underlying assumption here is that the dataset is large to the extent that  enumerating all possible data slices and validating model performance for each is not practical due to the sheer number of possible slices. Worse yet, simply searching for the most underperforming slices can be misleading because the model performance on smaller slices can be noisy, and without any safeguard, this leads to slices that are too small for meaningful impact on the model quality or that are false discoveries (i.e., non-problematic slices appearing as problematic). Ideally, we want to identify the largest and true problematic slices from the smaller slices that are not fully reflected on by the overall model performance metric.

%On the other hand, we also have more generic clustering-based algorithms \cite{krishnan2017palm,ribeiro2016explaining,anchors:aaai18} that group similar examples together as clusters and analyze model behavior locally within each cluster. 
There are more generic clustering-based algorithms in model understanding\cite{krishnan2017palm,ribeiro2016explaining,anchors:aaai18} that group similar examples together as clusters and analyze model behavior locally within each cluster. Similarly, we can cluster similar examples and treat each cluster as an arbitrary data slice; if a model underperforms on any of the slices, then the user can analyze the examples within.
However, clusters of similar examples can still have high variance and high cardinality of feature values, which are hard to summarize and interpret. In comparison, a data slice with a few common feature values (e.g., the {\sf Female} slice contains all examples with \predicate{Sex}{Female}) is much easier to interpret. In practice, validating and reporting model performance on interpretable slices are much more useful than validating on arbitrary non-interpretable slices (e.g., a cluster of similar examples with mixed properties).

A good technique to detect problematic slices for model validation thus needs to find easy-to-understand subsets of data and ensure that the model  performance on the subsets is meaningful and not attributed to chance.
Each problematic slice should be immediately understandable to a human without the guesswork.
The problematic slices should also be large enough so that their impact on the overall model quality is non-negligible.
Since the model may have a high variance in its prediction quality, we also need to be careful not to choose slices that are false discoveries.
Finally, since the slices have an exponentially large search space, it is infeasible to manually go though each slice. Instead, we would like to guide the user to a handful of slices that satisfy the conditions above. In this paper we propose \slicefinder{}, which efficiently discovers large possibly-overlapping slices that are both interpretable and problematic.

A slice is defined as a conjunction of feature-value pairs where having fewer pairs is considered more interpretable.
A problematic slice is identified based on testing of a significant difference of model performance metrics (e.g., loss function) of the slice and its counterpart. That is, we treat each problematic slice as a hypothesis and check that the difference is statistically significant, and the magnitude of the difference is large enough according to the effect size.
We discuss the details in  Section~\ref{sec:problem}.
One problem with performing many statistical tests (due to a large number of candidate slices) is an increased number of false positives. 
This is what is also known as Multiple Comparisons Problem (MCP) \cite{benjamini1995controlling}: imagine a test of Type-I error (false positive: recommending a non-problematic slice as problematic) rate of 0.05 (a common $\alpha$-level for statistical significance testing); the probability of having any false positives blows up exponentially with the number of comparisons (e.g., $1-(1-0.05)^8 = 0.34$, even for just 8 tests, but then, we may end up exploring hundreds and thousands of slices even for a modest number of examples). We address this issue in Section~\ref{sec:multi_hyp}.

In addition to testing, the slices found by \slicefinder{} can be used to evaluate model fairness or in applications such as fraud detection, business analytics, and anomaly detection, to name a few. While there are many definitions for fairness, a common one is that a model performs poorly (e.g., lower accuracy) on certain sensitive features (which define the slices), but not on others.
Fraud detection also involves identifying classes of activities where a model is not performing as well as it previously did. For example, some fraudsters may have gamed the system with unauthorized transactions. 
In business analytics, finding the most promising marketing cohorts can be viewed as a data slicing problem.
Although \slicefinder{} evaluates each slice based on its losses on a model, we can also generalize the data slicing problem where we assume a general scoring function to assess the significance of a slice. For example, data validation is the process of identifying training or validation examples that contain errors (e.g., values are out of range, features are missing, and so on). By scoring each slice based on the number or type of errors it contains, we can summarize the data errors through a few interpretable slices rather than showing users an exhaustive list of all erroneous examples.

The main contribution of this paper is applying data management techniques to the model validation problem in machine learning. This application is part of a larger integration of the areas of Big data and Artificial Intelligence (AI) where data management plays a role in almost all aspects of machine learning~\cite{polyzotis2018data,polyzotis2017data}. This paper extends our previous work on \slicefinder{}~\cite{chung2019slice,chung2018slice}. In particular, we provide a full description of the slice finding algorithms and provide extensive experiments.

In summary, we make the following contributions:
\squishlist %\begin{itemize}
\item We define the data slicing problem and the use of hypothesis testing for problematic slice identification (Section~\ref{sec:problem}) and false discovery control (Section~\ref{sec:multi_hyp}).
\item We describe the \slicefinder{} system and propose three automated data slicing approaches, including a na\"ive clustering-based approach as a baseline for automated data slicing (Section~\ref{sec:slicefinder}).
%\begin{itemize}
%    \item Decision Tree: Training a decision tree to partition the data into problematic and non-problematic slices.
%    \item Lattice: Searching a lattice of possibly overlapping slices.
%\end{itemize}
\item We present model fairness as a potential use case for \slicefinder{} (Section~\ref{sec:usecases}).
\item We evaluate the automated data slicing approaches using real and synthetic datasets (Section~\ref{sec:experiments}).
\squishend %\end{itemize}

\section{Data Slicing Problem}
\label{sec:problem}
\subsection{Preliminaries}
We assume a dataset
$D$ with $n$ examples and a model $h$ that needs to be tested. Following common practice, we assume that each example $x^{(i)}_F$ contains features  $F=\{F_1, F_2, ..., F_m\}$ where each feature $F_j$ (e.g., {\sf country}) has a list of values (e.g., \{{\sf US}, {\sf DE}\}) or discretized numeric value ranges (e.g., \{[0, 50), [50, 100)\}). We also have a ground truth label $y^{(i)}$ for each example, such that $D$ = $\{(x^{(1)}_F, y^{(1)})$, $(x^{(2)}_F, y^{(2)}), ... , (x^{(n)}_F, y^{(n)})\}$.
The test model $h$ is an arbitrary function that maps an input example to a prediction using $F$, and the goal is to validate if $h$ is working properly for different subsets of the data.  For ease of exposition,
we focus on a binary classification problem (e.g., UCI Census income classification) with $h$ that takes an example $x^{(i)}_F$ and outputs a prediction $h(x^{(i)}_F)$ of the true label $y^{(i)} \in \{0, 1\}$ (e.g., a person's income is above or below \$50,000).
Without loss of generality, we also assume that the model uses all the features in $F$ for classification.

A {\em slice} $S$ is a subset of examples in $D$ with common features and can be described as a predicate that is a conjunction of literals $\bigwedge_j F_j\ op\ v_j$ where the $F_j$'s are distinct (e.g., \predicate{country}{DE} $\wedge$ \predicate{gender}{Male}), and $op$ can be one of $=$, $\neq$, $<$, $\leq$, $\geq$, or $>$. For numeric features, we can discretize their values (e.g., quantiles or equi-height bins) and generate ranges so that they are effectively categorical features (e.g., \predicate{age}{[20,30)}). Numeric features with large domains tend to have fewer examples per value, and hence do not appear as significant. By discretizing numeric features into a set of continuous ranges, we can effectively avoid searching through tiny slices of minimal impact on model quality and group them to more sizable and meaningful slices.

We also assume a classification loss function $\psi(S, h)$ that returns a performance score for a set of examples by comparing $h$'s prediction $h(x^{(i)}_F)$ with the true label $y^{(i)}$. A common classification loss function is logarithmic loss (\emph{log loss}), which in case of binary classification is defined as:
\[
-\frac{1}{n} 
\sum_{(x^{(i)}_F,y^{(i)})\in S}[y^{(i)}\ \mathrm{ln}\ h(x^{(i)}_F) + (1 - y^{(i)})\ \mathrm{ln}\ (1 - h(x^{(i)}_F))]
\]
The log loss is non-negative and grows with the number of classification errors. A perfect classifier $h$ would have log loss of zero, and a random-guesser ($h(x)=0.5$) log loss of $-ln(0.5)=0.693$.
Also note that our techniques and the problem setup can easily generalize to other machine learning problem types (e.g., multi-class classification, regression, etc.) with proper loss functions/performance metrics.

\subsection{Model Validation}
\label{sec:slice_hypothesis}

We consider the model validation scenario of pointing the user to ``problematic'' slices where a single model performs relatively poorly on. That is, we would like to find slices where the loss function returns a significantly higher loss than the rest of the examples in $D$. At the same time, we prefer these slices to be large as well. For example, the slice \predicate{country}{DE} may be too large for a model to perform significantly worse than other countries. On the other hand, the slice \predicate{country}{DE} $\wedge$ \predicate{gender}{Male} $\wedge$ \predicate{age}{30} may have a high loss, but may also be too specific and thus small to have much impact on the overall performance of the model. Finally, we would like the slices to be interpretable in the sense that they can be expressed with a few literals. For example, \predicate{country}{DE} is more interpretable than \predicate{country}{DE} $\wedge$ \predicate{age}{20-40} $\wedge$ \predicate{zip}{12345}.

A straightforward extension of this scenario is to compare two models on the same data and point out if certain slices would experience a degrade in performance if the second model would be used. For example, a user may be using an existing model and wants to determine if a newly-trained model is safe to push to production. Here we can consider the two models as a single model where the loss is defined as the loss of the second model minus the loss of the first model. Since the extension does not fundamentally change the problem, for the rest of the paper, we focus on the original scenario of validating a single model.

Finding the most problematic slices is challenging because it requires a balance between how significant the difference in loss is and how large the slice is. Simply finding a slice with many classification errors will not work because there may also be many correct classifications within the same slice (recall that a slice is always of the form $\bigwedge_j F_j\ op\ v_j$). Another solution would be to score each slice based on some weighted sum of its size and difference in average losses. However, this weighting function is hard to tune by the user because it is not clear how size relates to loss. Instead, we envision the user to either fix the significance or size.

\subsection{Problematic Slice as Hypothesis}

We now discuss what we mean by significance in more detail. For each slice $S$, we define its counterpart $S'$ as $D-S$, which is the rest of the examples. We then compute the relative loss as the difference $\psi(S,h) - \psi(S',h)$. Without loss of generality, we only look for positive differences where the loss of $S$ is higher than that of $S'$. 

A key question is how to determine if a slice $S$ has a significantly higher loss than $S'$. Our solution is to treat each slice as a hypothesis and perform two tests: determine if the difference in loss is {\em statistically significant} and if the {\em effect size}~\cite{effectsize} of the difference is large enough. Using both tests is a common practice~\cite{doi:10.4300/JGME-D-12-00156.1} and necessary because statistical significance measures the existence of an effect (i.e., the slice indeed has a higher loss than its counterpart) while the effect size complements statistical significance by measuring the magnitude of the effect (i.e., how large the difference is).

To measure the statistical significance, we use the hypothesis testing with the following null ($H_o$) and alternative ($H_a$) hypotheses:
\[
H_o: \psi(S,h) \leq \psi(S',h)
\]
\[
H_a: \psi(S,h) > \psi(S',h)
\]
Here both $S$ and $S'$ should be viewed as samples of all the possible examples in the world, including the training data and even the examples that the model might serve in the future. We then use Welch's $t$-test~\cite{welch}, which is used to test the hypothesis that two populations have equal means and is defined as follows:
\[
t = \frac{\mu_S - \mu_{S'}}{\sqrt{\sigma_S^2 / |S| + \sigma_{S'}^2 / |S'|}}
\]
where $\mu_S$ is the average loss of $S$, $\sigma_S$ is the variance of the individual example losses in $S$, and $|S|$ is the size of $S$. In comparison to Student's $t$-test, Welch's $t$-test is more reliable when the two samples have unequal variances and unequal sample sizes, which fits our setting.

To measure the magnitude of the difference between the distributions of losses of $S$ and $S'$, we compute the effect size~\cite{effectsize} $\phi$, which is defined as follows:
\[
\phi = \sqrt{2} \times \frac{\psi(S,h) - \psi(S',h)}{\sqrt{\sigma_S^2 + \sigma_{S'}^2}}
\]

Intuitively, if the effect size is 1.0, we know that the two distributions differ by one standard deviation. According to Cohen's rule of thumb~\cite{cohen1988statistical}, an effect size of 0.2 is considered small, 0.5 is medium, 0.8 is large, and 1.3 is very large.

\subsection{Problem Definition}
\label{sec:data_slicing_problem}

For two slices $S$ and $S'$, we say that $S \prec S'$ if $S$ precedes $S'$ when ordering the slices by increasing number of literals, decreasing slice size, and decreasing effect size. Then the goal of \slicefinder{} is to identify {\em problematic slices} as follows:
\begin{definition}
\label{def:dataslicing}
Given a positive integer $k$, an effect size threshold $T$, and a significance level $\alpha$, find the top-$k$ slices sorted by the ordering $\prec$ such that:
\squishlist %\begin{itemize}
    \item [(a)] Each slice $S$ has an effect size at least $T$,
    \item [(b)] The slice is statistically significant,
    %\yeounoh{No slice can be replaced with any of its ancestors with fewer predicates and satisfy the above two conditions.}
    \item [(c)] No slice can be replaced with one that has a strict subset of literals and satisfies the above two conditions.
\squishend %\end{itemize}
\end{definition}

The top-$k$ slices do not have to be distinct, e.g., \predicate{country}{DE} and \predicate{education}{Bachelors} overlap in the demographic of Germany with a Bachelors degree. In a user's point of view, setting the effect size threshold $T$ may be challenging, so \slicefinder{} provides a slider for $T$ that can be used to explore slices with different degrees of problematicness (see Section~\ref{sec:interactive_viz}).

\section{System Architecture}
\label{sec:slicefinder}

\begin{figure}[t]
   \centering
     \includegraphics[width=\columnwidth]{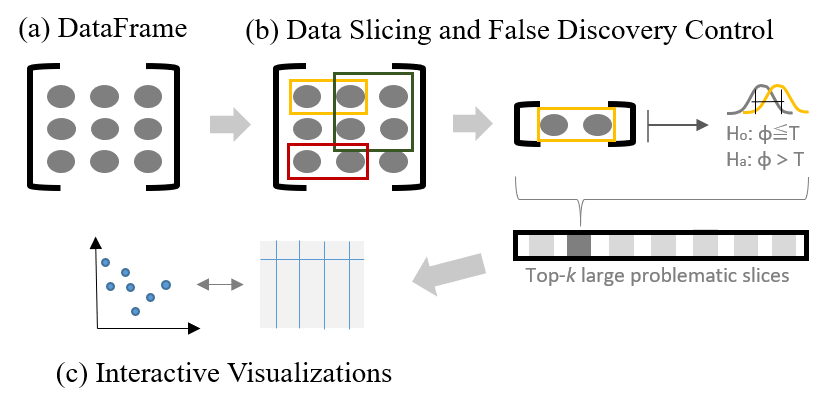}
     \caption{The \slicefinder{} architecture: (a) Data is loaded into a Pandas DataFrame, (b) \slicefinder{} performs automated data slicing and false discovery control to find the top-$k$ large problematic slices, and (c) \slicefinder{} provides interactive visualizations.}
 \label{fig:architecture}
\end{figure}

Underlying the \slicefinder{} system is an extensible architecture (Figure~\ref{fig:architecture}) that combines automated data slicing and interactive visualization tools. 
\slicefinder{} loads the validation data set into a Pandas DataFrame~\cite{mckinney2011pandas}. The DataFrame supports indexing individual examples, and each data slice keeps a subset of indices instead of a copy of the actual data examples. \slicefinder{} provides basic slice operators (e.g., intersect) based on the indices; only when evaluating the machine learning model on a given slice does \slicefinder{} access the actual data by the indices to test the model.
The Pandas library also provides a number of options to deal with dirty data and missing values. For example, one can drop \emph{NaN} (missing values) or any values that deviate from the column types as necessary. 

Once the data is loaded into a DataFrame, \slicefinder{} processes it to identify the problematic slices and allows the user to explore them. \slicefinder{} searches for problematic slices either by training a CART decision tree around misclassified examples or by performing a more exhaustive search on a lattice of slices. Both search strategies progress in a top-down manner until they find the top-$k$ problematic slices. The decision tree approach materializes the tree model and traverses the tree to find problematic slices. In lattice searching, \slicefinder{} traverses a lattice of slices to find the slices. This top-down approach allows \slicefinder{} to quickly respond to new request queries that use different $k$, as described in Section~\ref{sec:interactive_viz}. As \slicefinder{} searches through a large number of slices, some slices might appear problematic by chance (i.e., multiple comparisons problem \cite{foster2008a}). \slicefinder{} controls such a risk by applying a marginal false discovery rate (mFDR) controlling procedure called $\alpha$-investing \cite{foster2008a,DBLP:conf/sigmod/ZhaoSZBUK17} in order to find statistically significant slices among a stream of slices. Lastly, even a handful of problematic slices can be overwhelming to the user, since she may need to take action (e.g., deeper analyses or model debugging) on each slice. Hence, it is important to enable the user to quickly browse through the slices by slice size and effect size. To this end, \slicefinder{} provides interactive visualization tools for the user to explore the recommended slices.

The following subsections describe the \slicefinder{} components in detail. Section~\ref{sec:automaticdataslicing} introduces the automated data slicing approaches without false discovery control, Section~\ref{sec:multi_hyp} discusses the false discovery control, and Section~\ref{sec:interactive_viz} describes the interactive visualization.

\subsection{Automated Data Slicing}
\label{sec:automaticdataslicing}

As mentioned earlier, the goal of this component is to automatically identify problematic slices for model validation. 
To motivate the development of the two techniques that we mentioned (decision tree and lattice searching), let us first consider a simple baseline approach that identifies the problematic slices through clustering. And then, we discuss two automated data slicing approaches used in \slicefinder{} that improve on the clustering approach.

\subsubsection{Clustering}
The idea is to cluster similar examples together and take each cluster as an arbitrary data slice. If a test model fails on any of the slices, then the user can examine the data examples within or run a more complex analysis to fix the problem. This is an intuitive way to understand the model and its behavior (e.g., predictions) \cite{krishnan2017palm,ribeiro2016explaining,anchors:aaai18}; we can take a similar approach to the automated data slicing problem. The hope is that similar examples would  behave similarly even in terms of data or model issues. 

Clustering is a reasonable baseline due to its ease of use, but it has major drawbacks: first, it is hard to cluster and explain high dimensional data. 
We can reduce the dimensionality using principled component analysis (PCA) before clustering, but many features of clustered examples (in its original feature vector) still have high variance or high cardinality of values. Unlike an actual data slice filtered by certain features, this is hard to interpret unless the user can manually go through the examples and summarize the data in a meaningful way.
Second, the user has to specify the number of clusters, which affects crucially the quality of clusters in both metrics and size. As we want slices that are problematic and large (more impact for model quality), this is a key parameter, which is hard to tune. The two techniques that we present next overcome these deficiencies.

\subsubsection{Decision Tree Training}
To identify more interpretable problematic slices, we
train a decision tree that can classify which slices are problematic. The output is a partitioning of the examples into the slices defined by the tree.
For example, a decision tree could produce the slices \{$A > v$, $A \leq v \wedge B > w$, $A \leq v \wedge B \leq w$\}. For numeric features, this kind of partitioning is natural. For categorical features, a common approach is to use one-hot encoding where all possible values are mapped to columns, and the selected value results in the corresponding column to have a value 1. We can also directly handle categorical features by splitting a node using tests of the form $A = v$ and $A \neq v$.

%If a random forest is trained, then the one-hot encoding is not necessary.

To use a decision tree, we start from the root slice (i.e., the entire dataset) and go down the decision tree to find the top-$k$ problematic slices in a breadth-first traversal. The decision tree can be expanded one level at a time where each leaf node is split into two children that minimize impurity. The slices of each level are sorted by the $\prec$ ordering and then filtered based on whether they have large-enough effect sizes and are statistically significant. The details of the filtering are similar to lattice searching, which we describe in Section~\ref{sec:latticesearching}. The searching terminates when either $k$ slices are found or there are no more slices to explore.

The decision tree approach has the advantage that it has a natural interpretation, since the leaves directly correspond to slices. In addition, if the decision tree only needs to be expanded a few levels to find the top-$k$ problematic slices, then the slice searching can be efficient. On the other hand, the decision tree approach optimizes on the classification results and may not find all problematic slices according to Definition~\ref{def:dataslicing}. For example, if some feature is split on the root node, then it will be difficult to find single-feature slices for other features. In addition, a decision tree always partitions the data, so even if there are two problematic slices that overlap, at most one of them will be found. Another downside is that, if a decision tree gets too deep with many levels, then it starts to become uninterpretable as well~\cite{Freitas:2014:CCM:2594473.2594475}.

\subsubsection{Lattice Searching}
\label{sec:latticesearching}

\begin{figure}[t]
   \centering
     \includegraphics[width=\columnwidth]{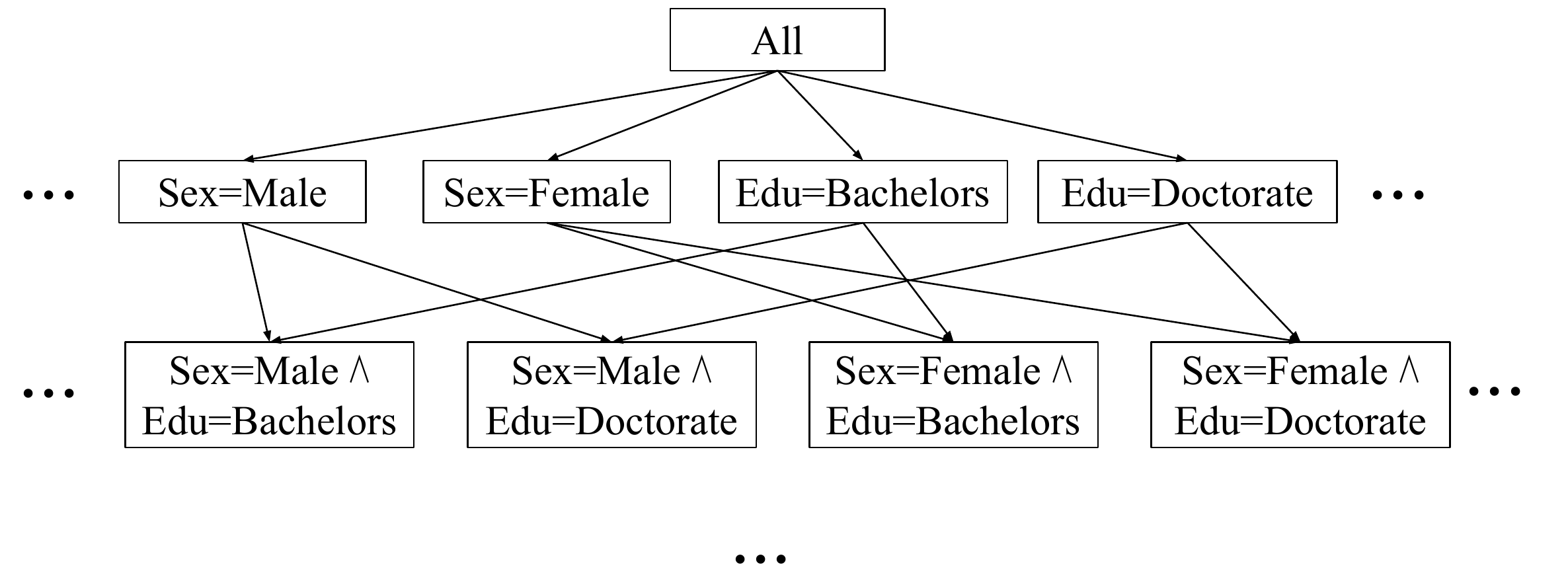}
     \caption{A lattice hierarchy of slices. In contrast with a decision tree, the search is more exhaustive and covers all possible feature combinations.}
 \label{fig:lattice}
\end{figure}

The lattice searching approach considers a larger search space where the slices form a lattice, and problematic slices can overlap with one another. 
%It is possible to merge slices to support other $op$s, but we leave this as future work. 
We assume that slices only have equality literals, i.e., $\bigwedge_i F_i = v_i$. In contrast to the decision tree training approach, lattice searching can be more expensive because it searches overlapping slices.

Figure~\ref{fig:lattice} illustrates how slices can be organized as a lattice. Lattice searching performs a breadth-first search and efficiently identifies problematic slices as shown in Algorithm~\ref{alg:lattice_search}. As a pre-processing step, \slicefinder{} takes the training data and discretizes numeric features. For categorical features that contain too many values (e.g., IDs are unique for each example), \slicefinder{} uses a heuristic where it considers up to the $N$ most frequent values and places the rest into an ``other values'' bucket. The possible slices of these features form a lattice where a slice $S$ is a parent of every $S$ with one more literal.

\IncMargin{1em}
\begin{algorithm}[t!]
\DontPrintSemicolon
\SetAlgoLined
\SetKwInOut{Input}{Input}\SetKwInOut{Output}{Output}
%\Indm
\Input{Lattice root $R$, max. number of slices to return $k$, effect size threshold $T$, significance level $\alpha$}
\Output{Problematic slices $S$}
%\Indp
$S$ = []\;
$C$ = $PriorityQueue$() \tcc*[f]{candidates for significance testing sorted by $\prec$}\;
$N$ = [] \tcc*[f]{non-problematic slices}\;
$L$ = 1 \tcc*[f]{number of literals}\;
$E$ = $ExpandSlices$(\{$R$\}, $L$)\\
\tcc*[f]{expand root slice}\;
$W$ = $\alpha$ \tcc*[f]{initialize $\alpha$ wealth}\;
\While{True}{
  \For{$slice \in E$}{
    \If{$EffectSize$(slice) $\geq$ T}{$C$.push($slice$)\;}
    \Else{$N$.append($slice$)\;}
  }
  \While{$C$ not empty}{
    $slice$ = $C$.pop()\;
    \eIf{$IsSignificant$($slice$, $W$)}{
      $S$.append($slice$)\;
      \If{$|S| = k$}{
        return $S$\;
      }
      $W = UpdateWealth$($W$, 1)\;
    }{
      $N$.append($slice$)\;
      $W = UpdateWealth$($W$, 0)\;
    }
  }
  $L$ += 1\;
  $E$ = $ExpandSlices$($N$, $L$)\;
  \If{$E$ is empty}{
    break\;
  }
}
return $S$\;
\caption{Lattice Searching Algorithm}
\label{alg:lattice_search}
\end{algorithm}
\DecMargin{1em}

\slicefinder{} finds the top-$k$ interpretable and large problematic slices sorted by the $\prec$ order by traversing the slice lattice in a breadth-first manner, one level at a time. Initially \slicefinder{} considers the slices that are defined with one literal. For each slice, \slicefinder{} checks if it has an effect size at least $T$ (using the $EffectSize$ function) and adds it to the priority queue $C$, which contains candidate slices that are sorted by the $\prec$ order. Next, \slicefinder{} pops slices from $C$ and tests for statistical significance using the $IsSignificant$ function. The testing can be done using $\alpha$-investing, which we discuss in Section~\ref{sec:multi_hyp}. Sorting the slices in the middle of the process using $C$ is important for the $\alpha$-investing policy used by \slicefinder{} as we explain later.

Each slice that has both a large enough effect size and is statistically significant is added to $N$ and later expanded using the $ExpandSlices$ function where we generate each new slice by adding a literal, only if the resulting slice is not subsumed by a previously-identified problematic slice. The intuition is that any subsumed (expanded) slice contains a subset of the examples of its parent and is smaller with more filter predicates (less interpretable); thus, we do not expand larger and already problematic slices. By starting from the slices whose predicates are single literals and expanding only non-problematic slices with one additional literal at a time (i.e., top-down search from lower order slices to higher order slices), we can generate a superset of all candidate slices. Depending on whether each slice satisfies the two conditions, \slicefinder{} updates the $\alpha$-wealth accordingly using the $UpdateWealth$ function (details on the updating strategy are discussed in Section~\ref{sec:multi_hyp}).

%\paragraph*{Example} 
\begin{example}
Suppose there are three features $A$, $B$, and $C$ with the possible values \{$a_1$\}, \{$b_1, b_2$\}, and \{$c_1$\}, respectively. Also say $k$ = 2, and the effect size threshold is $T$. Initially, the root slice is expanded to the slices \predicate{$A$}{$a_1$}, \predicate{$B$}{$b_1$}, \predicate{$B$}{$b_2$}, and \predicate{$C$}{$c_1$}, which are inserted into $E$. Among them, suppose that only \predicate{$A$}{$a_1$} has an effect size at least $T$ while the others do not. Then \predicate{$A$}{$a_1$} is added to $C$ for significance testing while the rest are added to $N$. Next, \predicate{$A$}{$a_1$} is popped from $C$ and is tested for statistical significance. Suppose the slice is significant and is thus added to $S$. Since $C$ is now empty, the slices in $N$ are expanded to \predicate{$B$}{$b_1$} $\wedge$ \predicate{$C$}{$c_1$} and \predicate{$B$}{$b_2$} $\wedge$ \predicate{$C$}{$c_1$}, which are not subsumed by the problematic slice \predicate{$A$}{$a_1$}. If \predicate{$B$}{$b_1$} $\wedge$ \predicate{$C$}{$c_1$} is larger and has both an effect size at least $T$ and is statistically significant, then the final result is [\predicate{$A$}{$a_1$}, \predicate{$B$}{$b_1$} $\wedge$ \predicate{$C$}{$c_1$}].
\end{example}
The following theorem formalizes the correctness of this algorithm for the slice-identification problem. The proof is a straightforward proof-by-contradiction and is omitted.
%\yeounoh{*** comment3: we do not prove the correctness of other approaches, for we use existing clustering and decision tree algorithms. *** }
%\alkis{If this theorem does not hold for the decision tree approach then should we mention it explicitly in that section? I.e., that the decision tree cannot solve the problem as defined in section 2? -- Steven: Done.}

\begin{theorem}
The \slicefinder{} slices identified by Algorithm~\ref{alg:lattice_search} satisfy Definition~\ref{def:dataslicing}.
\end{theorem}

% \begin{proof}
% Since we only add slices that have effect sizes at least $T$ and are statistically significant to the priority queue, the first two conditions are satisfied trivially.
% %\yeounoh{Should we mention that the second condition is enforced later using mFDR techniques?}
% The third condition can be proven to hold using contradiction. Suppose a slice $S$ that is popped from the queue has a large enough effect size, but its ancestor $S'$ has not been added 
% % \yeounoh{but its parent $S'$ has not been added to the result with the effect size at least $T$.}
% %but there is another slice $S'$ that has not yet been added to the result, but has the same size with fewer features and should have been added to the result first. 
% However, the ancestors of this slice must have been all popped and expanded before $S$ was popped. In addition, since $S'$ has fewer features than $S$, it should have been placed before $S$ in the queue (hence the contradiction).
% \end{proof}

\subsubsection{Scalability}

\slicefinder{} optimizes its search by expanding the filter predicate by one literal at a time. Unfortunately, this strategy does not solve the scalability issue of the data slicing problem completely, and \slicefinder{} could still search through an exponential number of slices, especially for large high-dimensional data sets. To this end, \slicefinder{} also takes the following two approaches for speeding up search.

\noindent\textbf{Parallelization: }
For lattice searching, evaluating a given model on a large number of slices one-by-one (sequentially) can be very expensive. In particular, computing the effect sizes is the performance bottleneck.
So instead, \slicefinder{} can distribute effect size evaluation jobs (lines 8--12 in Algorithm~\ref{alg:lattice_search}) by keeping separate priority queues $E_d$ for the different number of literals $d$. The idea is that workers take slices from the current $E$ in a round-robin fashion and evaluate them asynchronously; the workers push slices that have high effect sizes to the $C$ priority queue (for hypothesis testing) as they finish evaluating the slices. The significance testing on the slices in $C$ can be done by a single worker because the slices have already been filtered by effect size, and the significance testing can be done efficiently. In addition, the added memory and communication overheads are negligible compared to the time for computing the effect sizes. If $C$ is empty, but $|S| < k$, \slicefinder{} moves onto the next queue $E_{d+1}$ and continues searching until $|S| = k$. 

For DT, our current implementation does not support parallel learning algorithms for constructing trees. However, there exist a number of highly parallelizable learning processes for decision trees \cite{srivastava1999parallel}, which \slicefinder{} could implement to make DT more scalable. 

\noindent\textbf{Sampling: } \slicefinder{} can also scale by running on a sample of the entire dataset. The runtime of \slicefinder{} is proportional to the sample size, assuming that the runtime for the test model is constant for each example. By taking a sample, however, we also run the risk of false positives (non-problematic slices that appear problematic) and false negatives (problematic slices that appear non-problematic or completely disappear from the sample) due to a decreased number of examples. Since we are interested in large slices that are more impactful to model quality, we can disregard smaller false negatives that may have disappeared from the sample. In Section~\ref{sec:scalability}, we show that even for small sample sizes, most of the problematic slices can still be found. In Section~\ref{sec:multi_hyp}, we perform significance testing to filter slices that falsely appear as problematic or non-problematic. 

\begin{figure*}[t]
\centering
\includegraphics[width=0.9\textwidth]{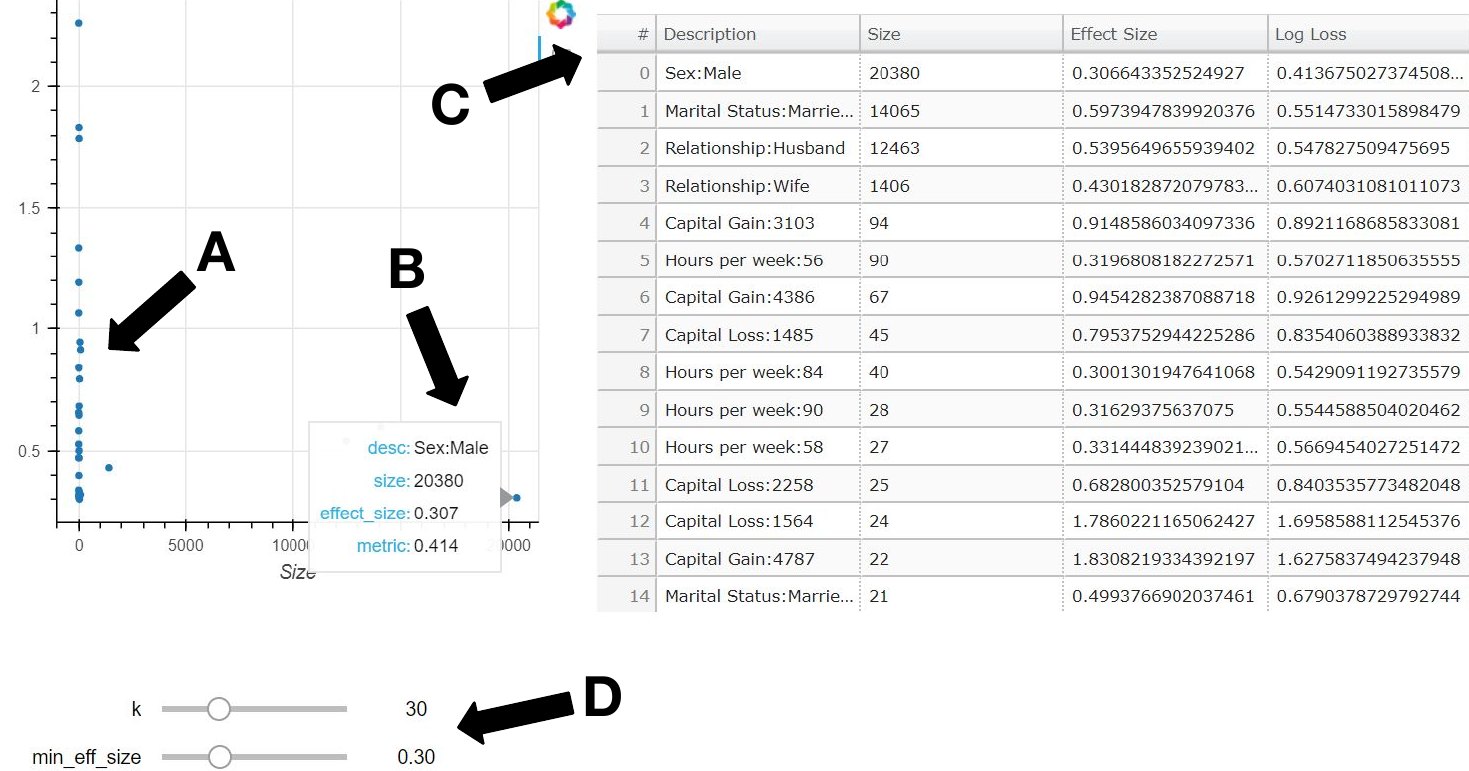}
\caption{The \slicefinder{} GUI helps users quickly browse through problematic slices by effect size and slice size on a scatter plot (A) and see a slice summary by hovering over any point (B); the user can sort slices by any metric and select on the scatter plot or table view. The selections are highlighted on the linked views (C). The user can also explore the top-$k$ large problematic slices by varying the effect size threshold using the slider (\emph{min\_eff\_size}) on the bottom left corner (D).}
\label{fig:app-slices}
\end{figure*}

\subsection{False Discovery Control}\label{sec:multi_hyp}

As \slicefinder{} finds more slices for testing, there is also the danger of finding more ``false positives'' (Type-1 errors), which are slices that are not statistically significant. \slicefinder{} controls false positives in a principled fashion using  $\alpha$-investing~\cite{foster2008a}. Given an $\alpha$-wealth (overall Type I error rate) $\alpha$, $\alpha$-investing spends this over multiple comparisons, while increasing the budget $\alpha$ towards the subsequent tests with each rejected hypothesis. This so called pay-out (increase in $\alpha$) helps the procedure become less conservative and puts more weight on more likely to be faulty null hypotheses. More specifically, an alpha-investing rule determines the wealth for the next test in a sequence of tests. This effectively controls marginal false discovery rate at level $\alpha$:
\[%\begin{equation}\label{eqn:alpha_guarantee}
\frac{\mathbb{E}(V)}{\mathbb{E}(R)} \leq \alpha
\]
Here, $V$ is the number of false discoveries and $R$ the number of total discoveries returned by the procedure.
\slicefinder{} uses $\alpha$-investing, mainly because it allows more interactive multiple hypothesis error control with an unspecified number of tests in any order. On the contrary, more restricted multiple hypothesis error control techniques, such as Bonferroni correction and Benjamini-Hochberg procedure \cite{benjamini1995controlling} fall short as they require the total number of tests $m$ in advance or become too conservative as $m$ grows large.

While there are different $\alpha$-investing policies~\cite{DBLP:conf/sigmod/ZhaoSZBUK17} for testing a sequence of hypotheses, we use a policy called {\em Best-foot-forward}. Recall our exploration strategy orders slices by decreasing slice size and effect size. As a result, the initial slices also tend to be statistically significant as well. The Best-foot-forward policy also assumes that many of the true discoveries are found early on and aggressively invests all $\alpha$-wealth on each hypothesis instead of saving some for subsequent hypotheses.

\subsection{Interactive Visualization Tool}
\label{sec:interactive_viz}

\slicefinder{} interacts with users through the GUI in Figure~\ref{fig:app-slices}. {\bf A}: On the left side is a scatter plot that shows the (size, effect size) coordinates of all slices. This gives an overview of the top-$k$ problematic slices, which allows the user to quickly browse through large and also problematic slices and compare slices to each other. {\bf B}: Whenever the user hovers a mouse over a dot, the slice description, size, effect size, and metric (e.g., log loss) are displayed next to it. If a set of slices are selected, their details appear on the table on the right-hand side, {\bf C}: On the table view, the user can sort slices by any metrics on the table. On the bottom, {\bf D}: \slicefinder{} provides  configurable sliders for adjusting $k$ and $T$. \slicefinder{} materializes all the problematic slices ($\phi \ge T$) as well as the non-problematic slices ($\phi < T$) searched already. If $T$ decreases, we just need to reiterate the slices explored until now to find the top-$k$ slices. If $T$ increases, then the current slices may not be sufficient, depending on $k$, so we continue searching the slice tree or lattice. This interaction is possible because \slicefinder{} looks for the top-$k$ slices in a top-down manner.

%\section{Case Study: Model Fairness}
\section{Using \slicefinder{} for Model Fairness}
\label{sec:usecases}
In this section, we look at model fairness as a potential use case of \slicefinder{} where identifying problematic slices can be a preprocessing step before more sophisticated analyses on fairness on the slices. As machine learning models are increasingly used in sensitive applications, such as predicting whether individuals will default on loans~\cite{hardt2016equality}, commit crime~\cite{machinebias}, or survive intensive hospital care~\cite{Ghassemi:2014:UPS:2623330.2623742}, it is essential to make sure the model performs equally well for all demographics to avoid discrimination.
However, models may fail this property for various reasons: bias in data collection, insufficient data for certain slices, limitations in the model training, to name a few cases.

%For sensitive attributes (e.g., gender), such difference can be viewed as discrimination.\alkis{I don't understand this last sentence and how it ties with the rest of the paragraph.}
%The bias can be resolved by augmenting the problematic data using techniques including crowdsourcing.

Model fairness has various definitions depending on the application and is thus non-trivial to formalize~\cite{barocas2017fairness}. While many metrics have been proposed~\cite{hardt2016equality,Dwork:2012:FTA:2090236.2090255,DBLP:conf/innovations/KleinbergMR17,Feldman:2015:CRD:2783258.2783311}, there is no widely-accepted standard, and some definitions are even at odds. 
%To illustrate how non-trivial it is to define fairness, Table~\ref{tbl:fairness-metrics} shows a subset of metrics from the literature.
In this paper, we focus on a relatively common definition, which is to find the data where the model performs relatively worse using some of these metrics, which fits into the \slicefinder{} framework.

Using our definition of fairness, \slicefinder{} can be used to quickly identify interpretable slices that have fairness issues without having to specify the sensitive features in advance. Here, we demonstrate how \slicefinder{} can be used to find any unfairness of the model with equalized odds~\cite{hardt2016equality}. Namely, we explain how our definition of a problematic slice using effect size also conforms to the definition of equalized odds. \slicefinder{} is also generic and supports any fairness metric that can be expressed as a scoring function. Any subsequent analysis of fairness on these slices can be done afterwards.

Equalized odds requires a predictor $\hat{Y}$ (e.g., a classification model $h$ in our case) to be independent of protected or sensitive feature values $A\in\{0,1\}$ (e.g., \predicate{gender}{Male} or \predicate{gender}{Female}) conditional on the true outcome $Y$ \cite{hardt2016equality}.
In binary classification ($ y \in \{0, 1\}$), this is equivalent to:
\[%\begin{equation}\label{eqn:equal_odds}
    Pr\{\hat{Y}=1 | A=0, Y=y\} = Pr\{ \hat{Y}=1 | A=1, Y=y\}
\]
Notice that equalized odds is essentially matching true positive rates (\emph{tpr}) in case of $y=1$ or false negative rates (\emph{fnr}) otherwise.
%Figure~\ref{fig:roc_gender} illustrates that the random forest classifier from Example~\ref{example_intro} is discriminatory with respect to gender because \emph{tpr} and \emph{fnr} do not match between the two gender demographics (one is a counterpart of the other).

%\begin{figure}[t]
%  \centering
%     %\includegraphics[width=0.5\textwidth, trim={8mm 15mm 2mm 2mm}, clip=true]{figure/clusters.png}
%     \includegraphics[width=0.9\columnwidth]{figure/roc_gender.png}
%     \caption{The Receiver Operating Characteristic curves for \predicate{gender}{Male} (effect size of $0.47$) and its counterpart. The gap between the two curves indicates the discriminatory behavior of the random forest classifier (from \emph{Example}~\ref{example_intro}) by equalized odds. Likewise, all slices with high effect size (e.g., $>T=0.4$) show such a discriminatory behavior.}
% \label{fig:roc_gender}
%\end{figure}

\slicefinder{} can be used to identify slices where the model is potentially discriminatory; a machine learning practitioner can easily identify feature dimensions of the data, without having to manually consider all feature value pair combinations, on which a deeper analysis and potential model fairness adjustments are needed.
The problematic slices with $\phi \geq T$ suffer from higher loss (lower model accuracy in case of log loss) compared to the counterparts.
If one group is enjoying a better rate of accuracy over the other, then it is a good indication that the model is biased. Namely, accuracy is a weighted sum of \emph{tpr} and \emph{fnr} by their proportions, and thus, a difference in accuracy means there are differences in \emph{tpr} and false positive rate ($fpr=1-tpr$), assuming there are any positive examples. As equalized odds requires matching \emph{tpr} and \emph{fpr} between the two demographics (a slice and its counterpart), \slicefinder{} using $\psi$ can identify slices to show that the model is potentially discriminatory.
In case of the \predicate{gender}{Male} slice above, we flag this as a signal for discriminatory model behavior because the slice is defined over a sensitive feature and has a high effect size.

\iffalse
We can correct the unfair  classifier for equalized odds by solving the following linear program for optimal $\theta = \{Pr\{\tilde{Y_\theta}=1 | A, \hat{Y}\}$ \cite{hardt2016equality} :
\begin{align*}
\text{min} \quad
\mathbb{E}\mathcal{L}(\tilde{Y_\theta}, Y)& \\
\text{s.t.} \quad
Pr\{\tilde{Y_\theta}=1 | A=1, Y=0\} &= Pr\{\tilde{Y_\theta}=1 | A=0, Y=0\} \\
Pr\{\tilde{Y_\theta}=1 | A=1, Y=1\} &= Pr\{\tilde{Y_\theta}=1 | A=0, Y=1\} \\
\forall_{a, y} 0 \leq Pr\{\tilde{Y_\theta}=1 & | A=a, Y=y\} \leq 1
\end{align*}
The derived classifier $\tilde{Y}$ predicts based on the protected feature $A$ and unfair predictions $\hat{Y}$ according to a solution $\theta$. Notice that the equality constraints enforces equalized odds.
\fi

There are other standards, but equalized odds ensures that the prediction is non-discriminatory with respect to a specified protected attribute (e.g., gender), without sacrificing the target utility (i.e., maximizing model performance) too much \cite{hardt2016equality}.

% \section{Other Applications}
% \label{sec:otherapplications}
% In Section~\ref{sec:usecases}, we describe a use case where we take a model and validate it with different subsets of a single data set for model fairness. In addition, \slicefinder{} can identify problematic slices in slightly different contexts.

% \paragraph*{Fraud Detection}
% Finding fraudulent users involves identifying demographics where a model is not performing as well as it previously did. For example, some fraudsters may have gamed the system with unauthorized transactions. Here \slicefinder{} can be used to identify potentially problematic slices for two datasets where the same model performs most differently. The slices can be used to narrow down the search space (e.g., users in a certain country of a certain age range have suspicious behavior) and then apply more sophisticated detection techniques on the slices.

% \paragraph*{Anomaly Detection}
% Although \slicefinder{} assumes a model, the data slicing problem can be more general where it uses any scoring function instead of a model for finding any anomalies in the data. For example, data validation is the process of identifying various data errors (e.g., values are out of range, features are not present in enough examples). Instead of showing users an exhaustive list of all errors, it can be useful to show the slices with many errors. A summary of anomalies can help find the root cause of the problem, which may appear as multiple symptoms.

\section{Experiments}
\label{sec:experiments}

In this section, we compare the two \slicefinder{} approaches (decision tree and lattice search) with the baseline (clustering). For the clustering approach, we use the $k$-means algorithm. We address the following key questions:
\squishlist % \begin{itemize}
    \item How accurate and efficient is \slicefinder{}?
    \item What are the trade-offs between the slicing techniques?
    \item What is the impact of adjusting the effect size threshold~$T$?
    \item Are the identified slices interpretable enough to understand the model's performance?
    \item How effective is false discovery control using $\alpha$-investing?
\squishend %\end{itemize}

\subsection{Experimental Setup}
\label{sec:setup}

We used the following two problems with different datasets and models to compare how the three different slicing techniques -- lattice search (LS), decision tree (DT), and clustering (CL) -- perform in terms of recommended slice quality as well as their interpretability.

\squishlist % \begin{itemize}
    \item {\em Census Income Classification: } We trained a random forest classifier (Example~\ref{example_intro}) to predict whether the income exceeds \$50K/yr based on the UCI census data \cite{kohavi1996scaling}. There are 15 features and 30K examples.
    \item {\em Credit Card Fraud Detection: } We trained a random forest classifier to predict fraudulent transactions among credit card transactions \cite{dal2015calibrating}. This dataset contains transactions that occurred over two days, where we have 492 frauds out of 284k transactions (examples), each with 29 features. Because the data set is heavily imbalanced, we first undersample non-fraudulent transactions to balance the data. This leaves a total of 984 transactions in the balanced dataset.
\squishend %\end{itemize}

As we shall see, the two datasets -- Census Income and Credit Card Fraud -- have different characteristics and are thus useful for comparing the behaviors of the decision tree and lattice search algorithms. In addition, we also use a synthetic dataset when necessary. The main advantage of using synthetic data is that it gives us more insights into the operations of \slicefinder{}. In Sections~\ref{sec:problematic}--\ref{sec:interpretability}, we assume that all slices are statistically significant for simplicity and separately evaluate statistical significance in Section~\ref{sec:falsediscoverycontrol}.

\noindent\textbf{Accuracy Measure: } Since problematic slices may overlap, we define {\em precision} to be the fraction of examples in the union of the slices identified by the algorithm being evaluated that also appear in actual problematic slices. Similarly, {\em recall} is defined as the fraction of the examples in the union of actual problematic slices that are also in the identified slices. Finally, {\em accuracy} is the harmonic mean of precision and recall.

\subsection{Problematic Slice Identification}
\label{sec:problematic}

An important question to answer is whether \slicefinder{} can indeed find the most problematic slices, in the user's point of view. Unfortunately for the real datasets, we do not know what are the true problematic slices, which makes our evaluation challenging. Instead, we {\em add new problematic slices} by randomly perturbing labels and focus on finding those slices. While \slicefinder{} may find both new and existing problematic slices, our evaluation will only be whether \slicefinder{} finds the new problematic slices.

We first experiment on a synthetic dataset and compare the performances of LS, DT, and CL. We then experiment on the real datasets and show similar results. 

\subsubsection{Synthetic Data}
\label{sec:synthetic_data}

We generate a simple synthetic dataset where the generated examples have two discretized features $F_1$ and $F_2$ and can be classified into two classes -- 0 and 1 -- perfectly. We make the model use this decision boundary and do not change it further. Then we add problematic slices by choosing random possibly-overlapping slices of the form $F_1 = A$, $F_2 = B$, or $F_1 = A \wedge F_2 = B$. For each slice, we flip the labels of the examples with 50\% probability. Note that this perturbation results in the worst model accuracy possible.

\begin{figure}[!t]
\centering

\begin{subfigure}{0.48\columnwidth}
     \includegraphics[trim={0 0 1cm 0},width=\columnwidth]{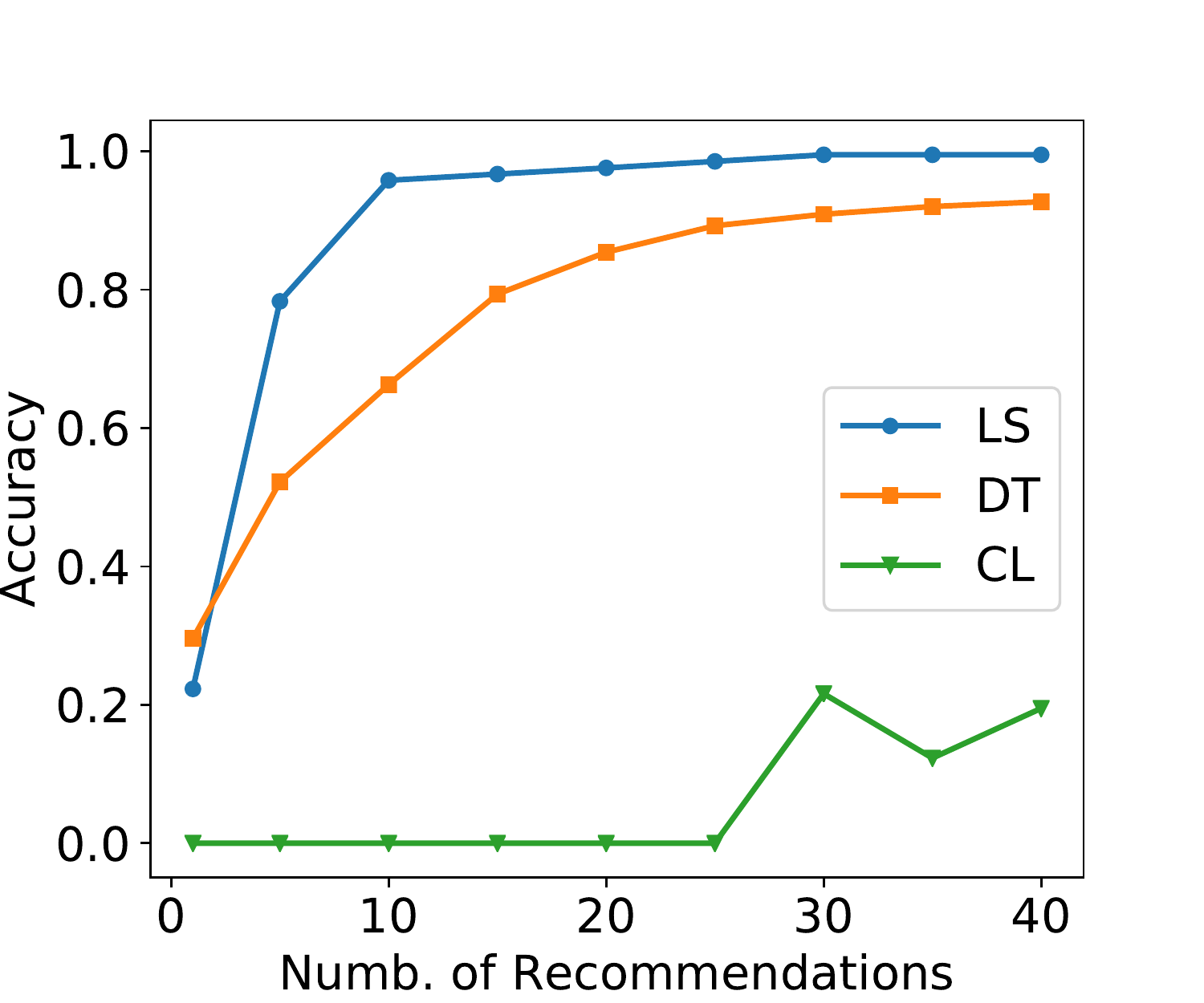}
     \caption{Synthetic Data}
\end{subfigure} 
\begin{subfigure}{0.48\columnwidth}
     \includegraphics[trim={0 0 1cm 0},width=\columnwidth]{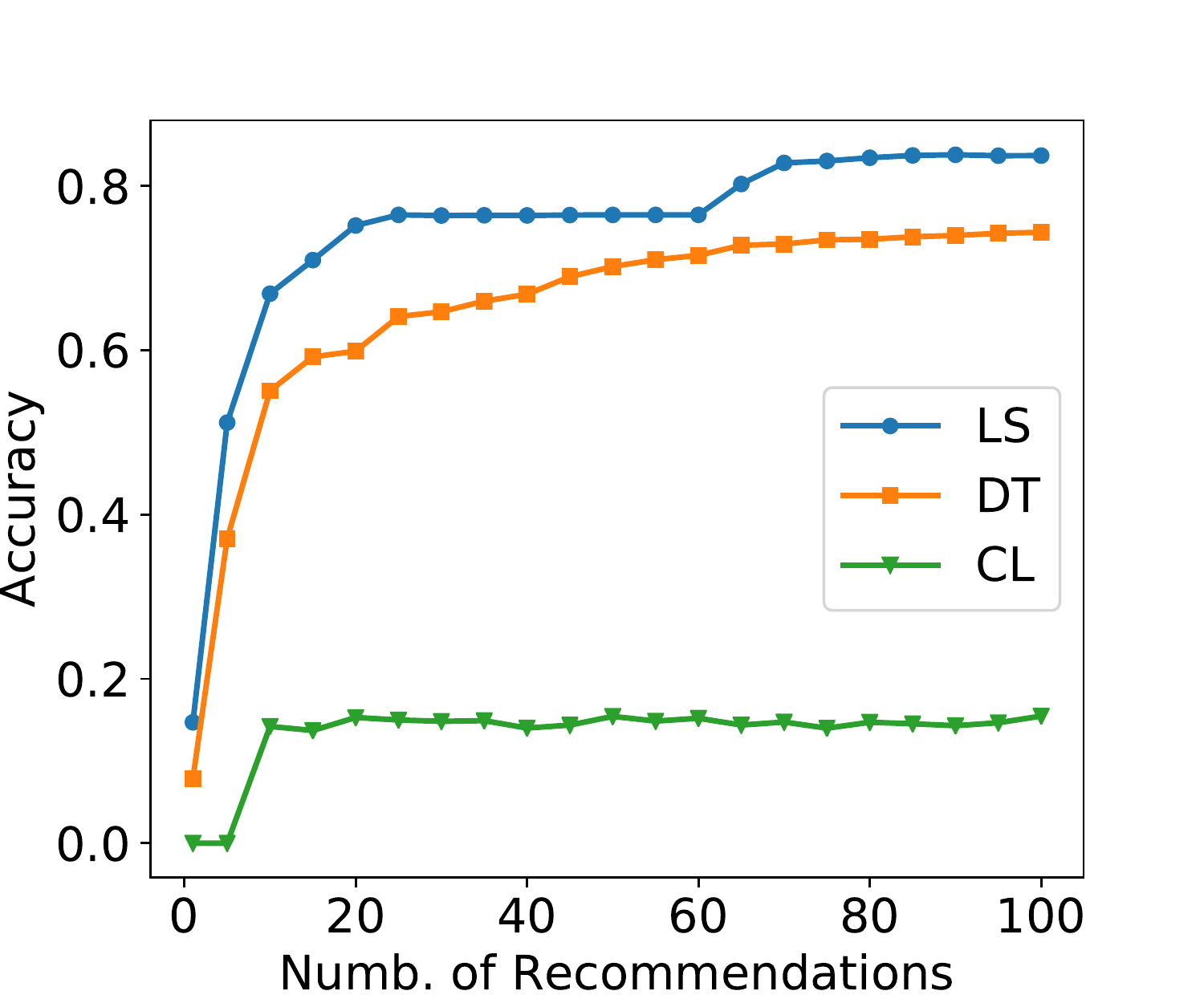}
     \caption{Census Income Data}
\end{subfigure}

\caption{Accuracy comparison of finding problematic slices using (a) synthetic data and (b) real data.}

\label{fig:problematic_slices}
\end{figure}

Figure~\ref{fig:problematic_slices}(a) shows the accuracy comparison of LS, DT, and CL on synthetic data. As the number of recommendations increases, LS consistently has a higher accuracy than DT because LS is able to better pinpoint the problematic slices including overlapping ones while DT is limited in the sense that it only searches non-overlapping slices. For CL, we only evaluated the clusters with effect sizes at least $T$. Even so, the accuracy is much lower than those of LS~and~DT.

\subsubsection{Real Data}

We also perform a similar experiment using the Census Income dataset where we generate new problematic slices on top of the existing data by randomly choosing slices and flipping labels with 50\% probability. Compared to the synthetic data, the existing data may also have problematic slices, which we do not evaluate because we do not know what they are. Figure~\ref{fig:problematic_slices}(b) shows similar comparison results between LS, DT, and CL. The accuracies of LS and DT are lower than those in the synthetic data experiments because some of the identified slices may be problematic slices in the existing data, but are considered incorrect when evaluated.

\subsection{Large Problematic Slices}

\begin{figure}[t!]
\centering
\begin{subfigure}{0.48\columnwidth}
     \includegraphics[width=\columnwidth]{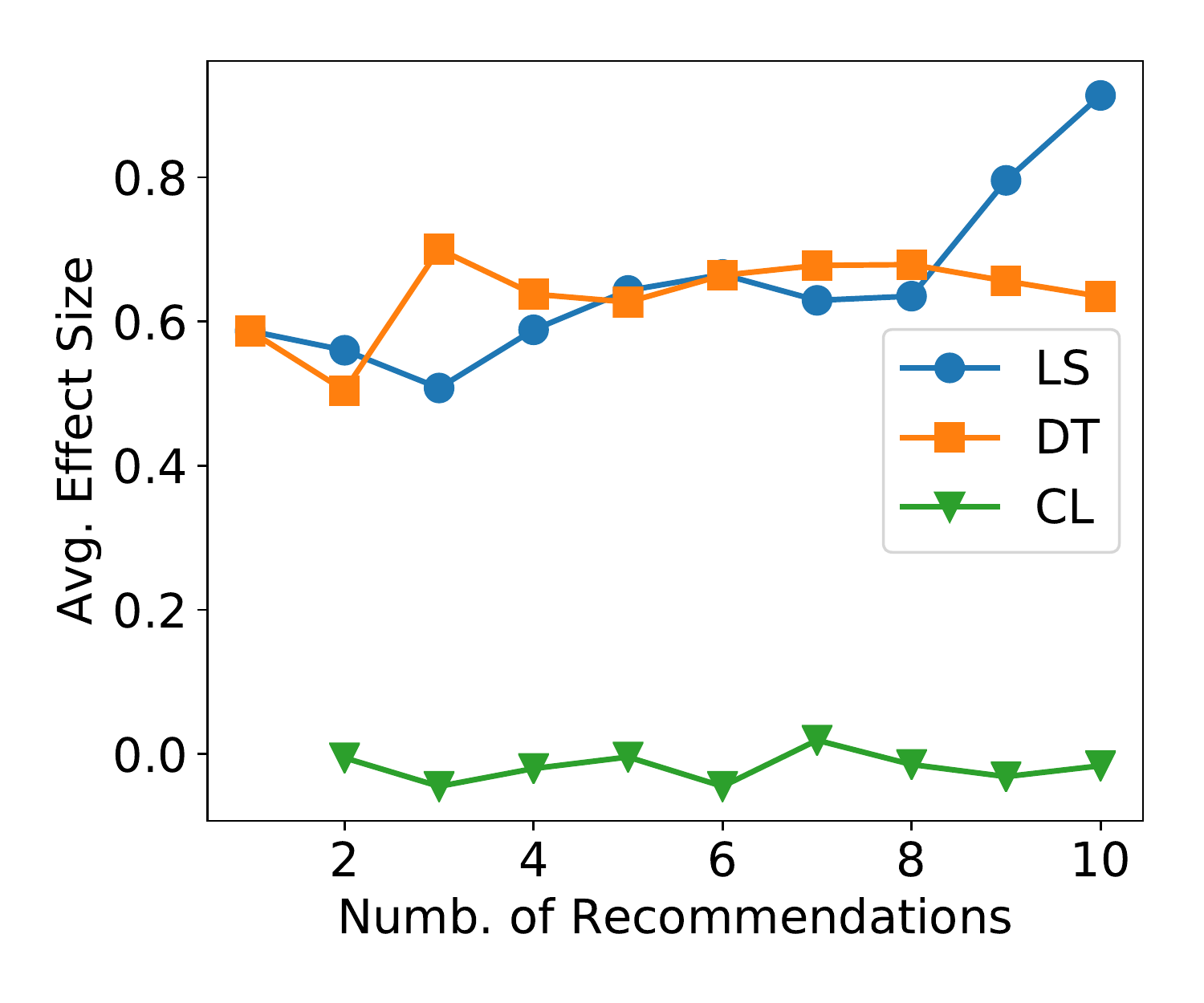}
     \caption{Census Income Data}
\end{subfigure}
\begin{subfigure}{0.48\columnwidth}
     \includegraphics[width=\columnwidth]{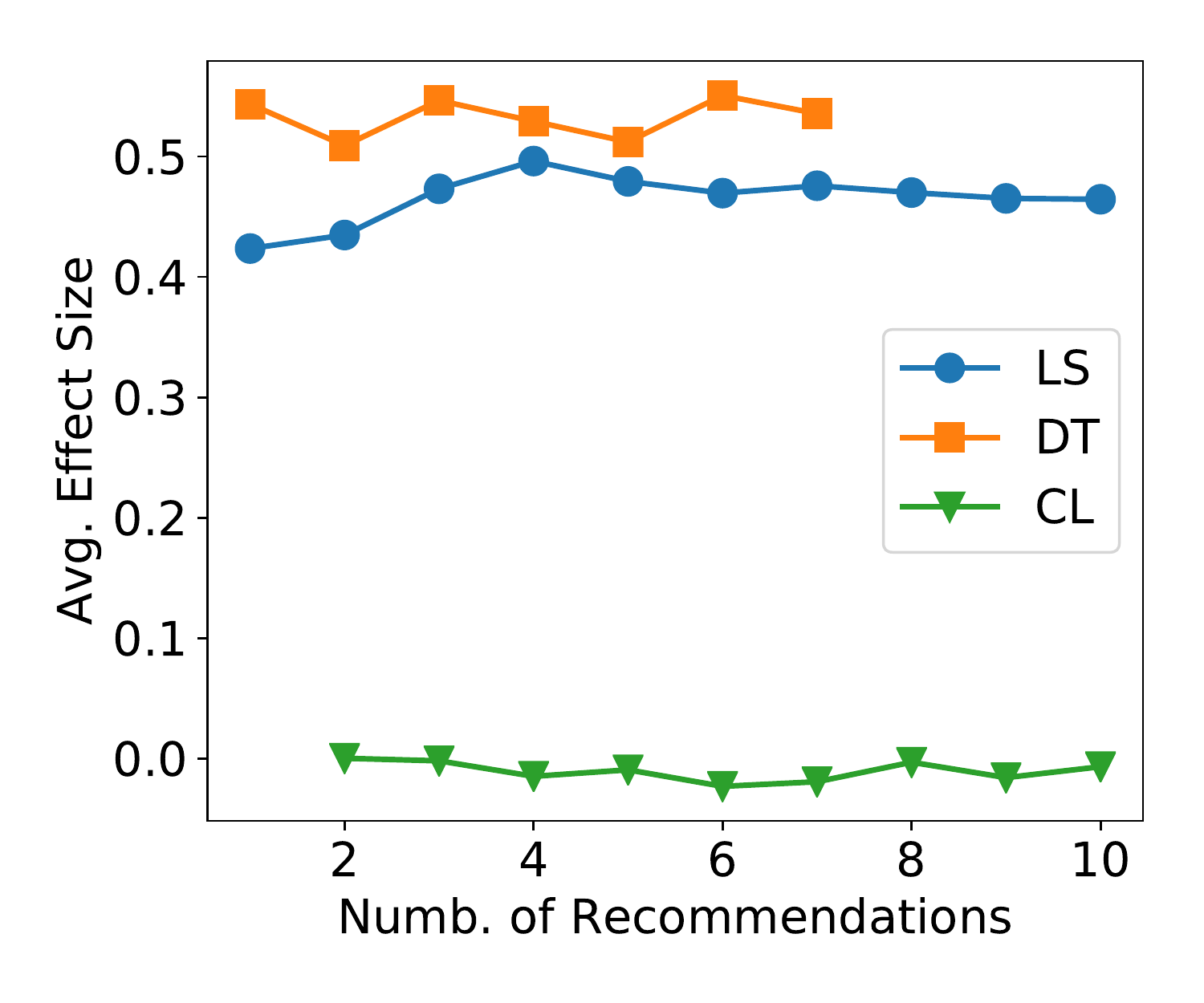}
     \caption{Credit Card Fraud Data}
\end{subfigure} 

\caption{Effect size comparisons between different data slicing approaches ($T = 0.4$).}
\label{fig:effect_size_comparison}
\end{figure}

\begin{figure}[t]
\centering
\begin{subfigure}{0.48\columnwidth}
     \includegraphics[width=\columnwidth]{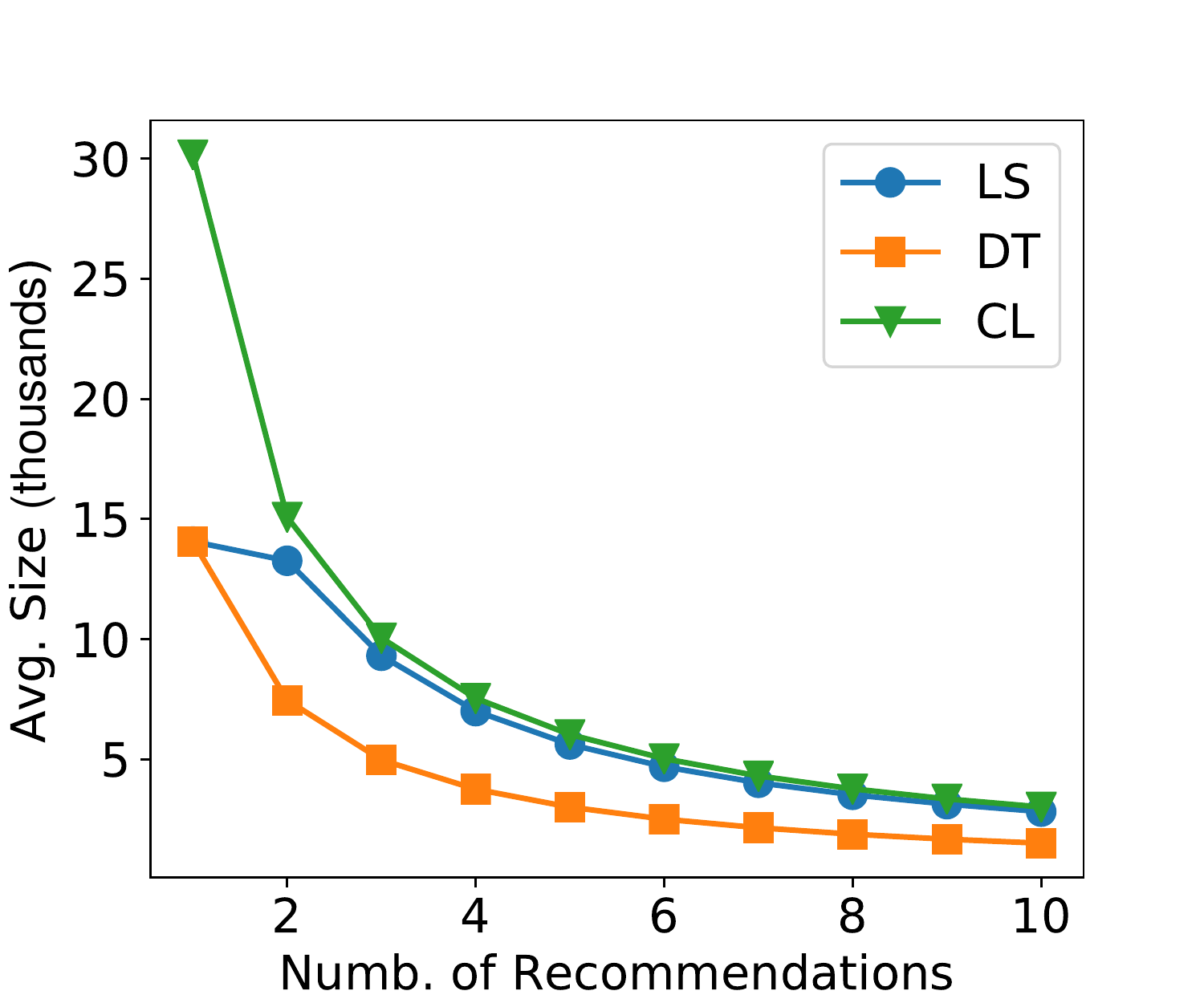}
     \caption{Census Income Data}
\end{subfigure}
\begin{subfigure}{0.48\columnwidth}
     \includegraphics[width=\columnwidth]{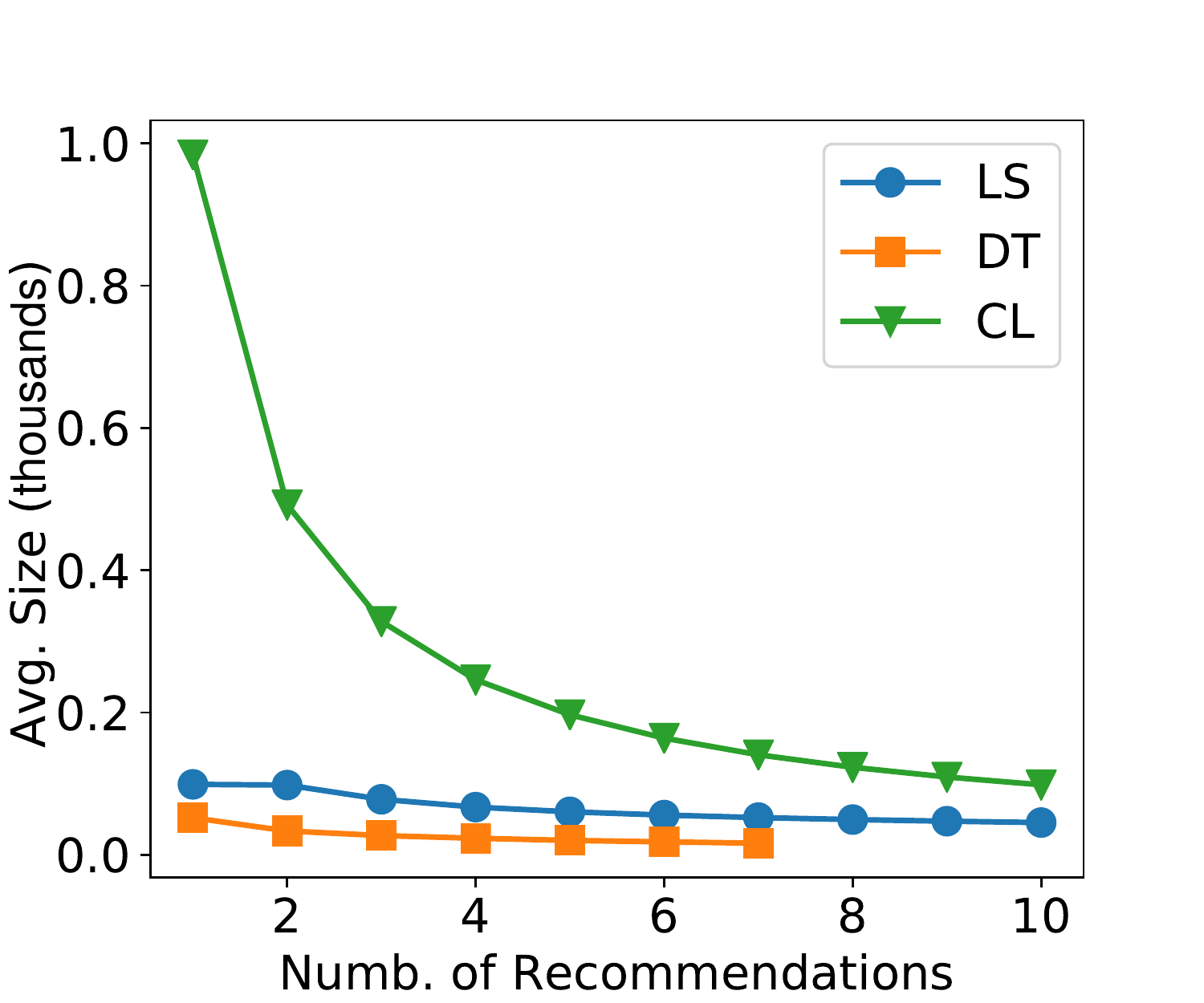}
     \caption{Credit Card Fraud Data}
\end{subfigure} 
\caption{Average slice size (unit is 1000) comparisons between different data slicing approaches ($T = 0.4$).}
\label{fig:size_comparison}
\end{figure}

Figures~\ref{fig:effect_size_comparison}~and~\ref{fig:size_comparison} show how LS and DT outperform CL in terms of average slice size and average effect size on the real datasets. CL starts with the entire dataset where the number of clusters (i.e., recommendations) is 1. CL produces large clusters that have very low effect sizes where the average is around 0.0 and sometimes even negative, which means some slices are not problematic. The CL results show that grouping similar examples does not necessarily guide users to problematic slices. In comparison, LS and DT find smaller slices with effect sizes above the threshold $T=0.4$.

LS and DT show different behaviors depending on the given dataset. When running on the Census Income data, both LS and DT are able to easily find up to $k=10$ problematic slices with similar effect sizes. Since LS generally has a larger search space than DT where it also considers overlapping slices, it is able to find larger slices as a result. When running on the Credit Card Fraud data, DT has a harder time finding enough problematic slices. The reason is that DT initially finds a large problematic slice, but then needs to generate many levels of the decision tree to find additional problematic slices because it only considers non-overlapping slices. Since a decision tree is designed to partition data to minimize impurity, the slices found deeper down the tree tend to be smaller and ``purer,'' which means the problematic ones have higher effect sizes. Lastly, DT could not find more than 7 problematic slices because the leaf nodes were too small to split further. These results show that, while DT may search a level of a decision tree faster than LS searching a level of a lattice, it may have to search more levels of the tree to make the same number of recommendations.

\subsection{Adjusting Effect Size Threshold $T$}

\begin{figure}[t]
\centering
\begin{subfigure}{\columnwidth}
     \includegraphics[width=\columnwidth]{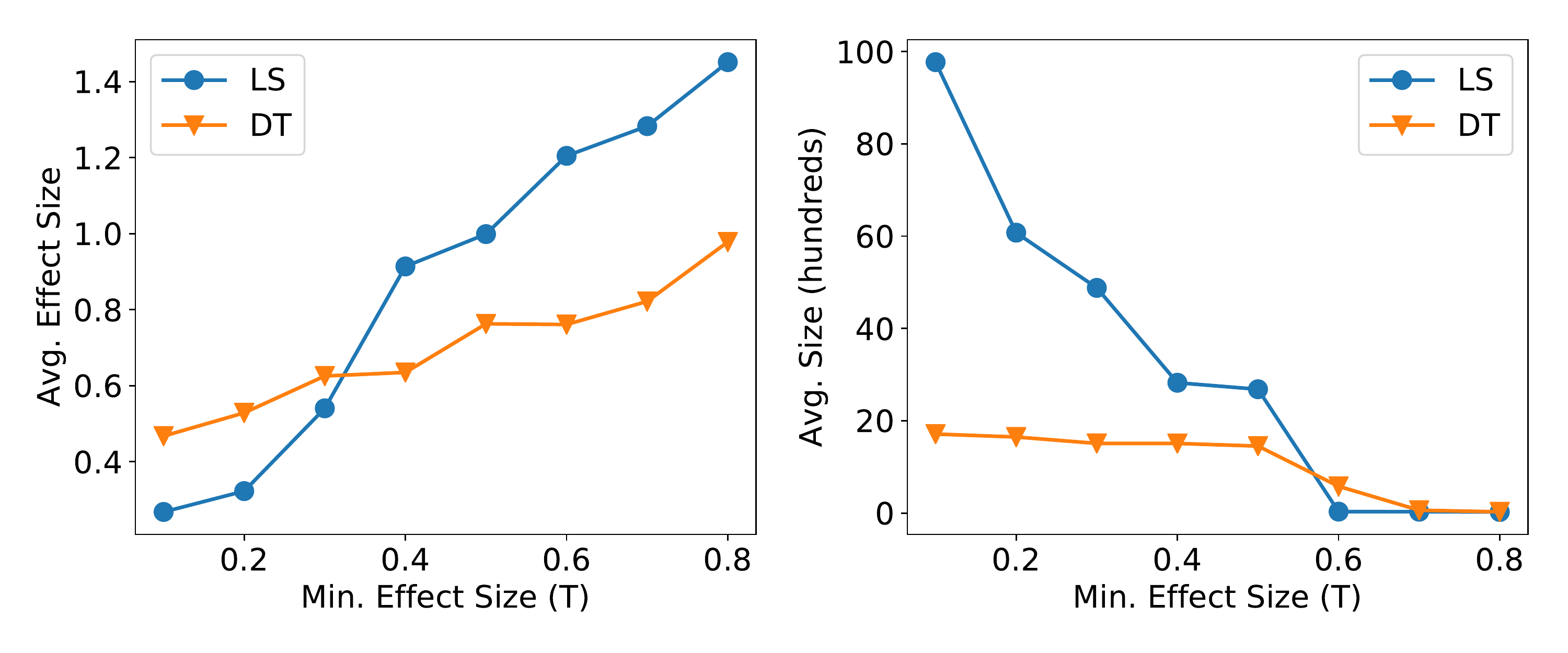}
     \caption{Census Income Data}
\end{subfigure}
\begin{subfigure}{\columnwidth}
     \includegraphics[width=\textwidth]{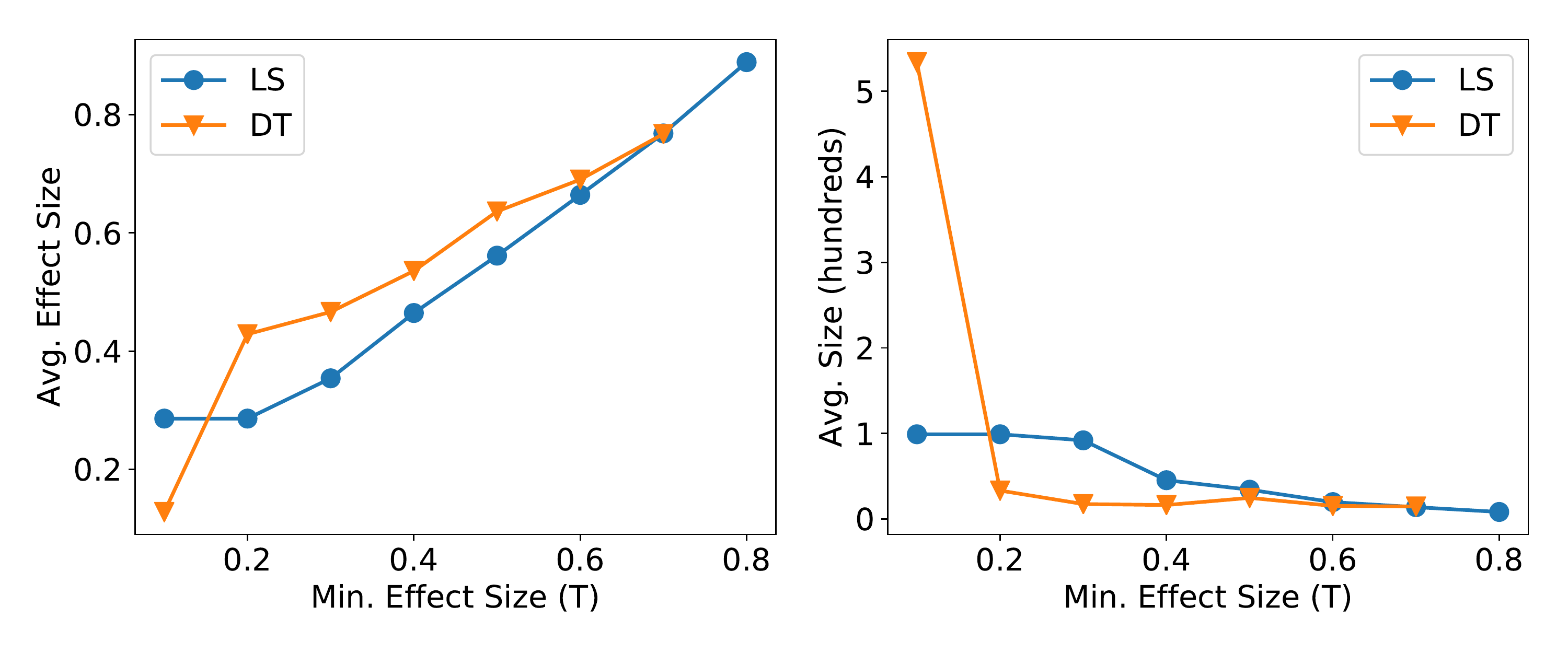}
     \caption{Credit Card Fraud Data}
\end{subfigure} 
\caption{The impact of adjusting the effect size threshold $T$ on average slice size and average effect size.}
\label{fig:slider_T}
\end{figure}

Figure~\ref{fig:slider_T} shows the impact of adjusting the effect size threshold $T$ on LS and DT. For a low $T$ value, there are more slices that can be problematic. Looking at the Census Income data, LS indeed finds larger slices than those found by DT, although they have relatively smaller effect sizes as a result. As $T$ increases, LS is forced to search smaller slices that have high-enough effect sizes. Since LS still has a higher search space than DT, it does find slices with higher effect sizes when $T$ is at least 0.4. The Credit Card Fraud data shows a rather different comparison. For small $T$ values, recall that DT initially finds a large problematic slice, which means the average size is high, and the effect size small. As $T$ increases, DT has to search many levels of the decision tree to find additional problematic slices. These additional slices are much smaller, which is why there is an abrupt drop in average slice size. However, the slices have higher effect sizes, which is why there is also a corresponding jump in the average effect size.

\subsection{Scalability}
\label{sec:scalability}

We evaluate the scalabilities of LS and DT against different sample fractions, degree of parallelization, and the number of top-$k$ slices to recommend. All experiments were performed on the Census Income dataset.

\begin{figure}[t]
\centering
\begin{subfigure}{0.49\columnwidth}
     \includegraphics[width=\columnwidth]{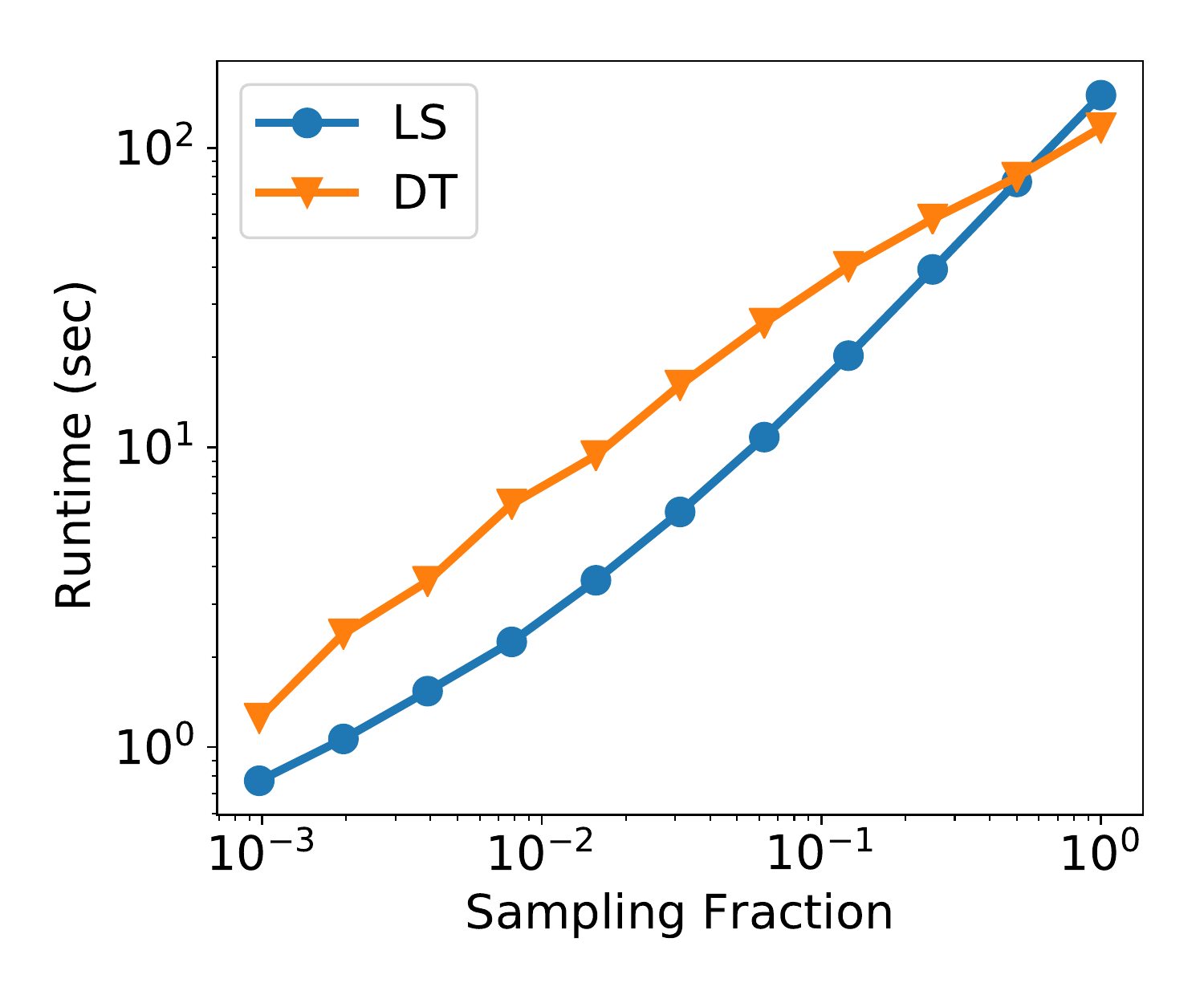}
\end{subfigure}
\begin{subfigure}{0.49\columnwidth}
     \includegraphics[width=\columnwidth]{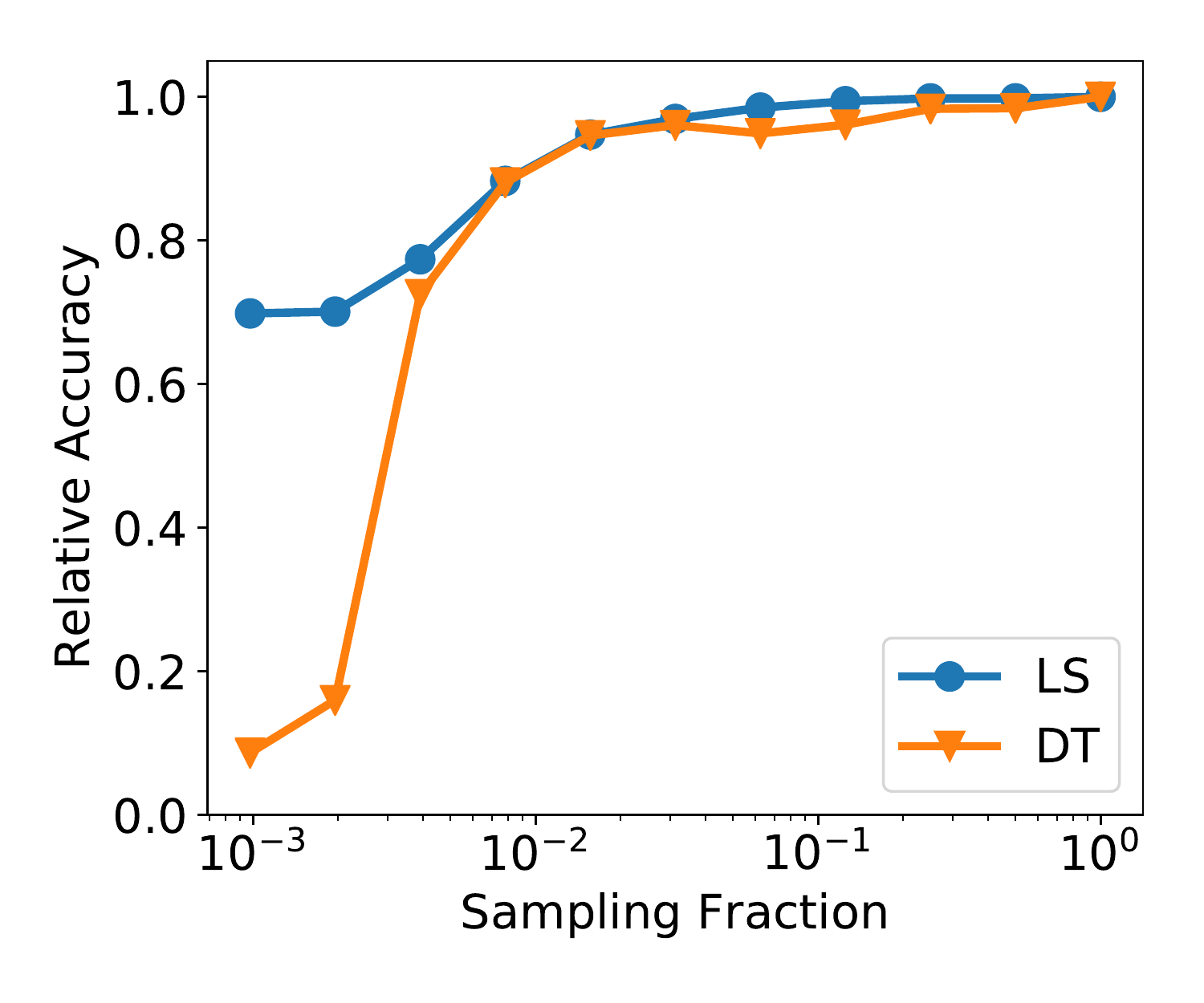}
\end{subfigure} 
\caption{\slicefinder{} (LS, DT) runtime (on a single node) and accuracy results using different sample sizes (Census Income data).}
\label{fig:scalability}
\end{figure}

Figure~\ref{fig:scalability} shows how the runtimes of LS and DT change versus the sampling fraction. For both algorithms, the runtime increases almost linearly with the sample size. We also measure the relative accuracy of the two algorithms where we compare the slices found in a sample with the slices found in the full dataset. For a sample fraction of 1/128, both LS and DT maintain a high relative accuracy of 0.88. These results show that it is possible to find most of the problematic slices using a fraction of the data, about two orders of magnitude faster.

\begin{figure}[t]
\centering
\begin{subfigure}{0.48\columnwidth}
\includegraphics[width=\columnwidth]{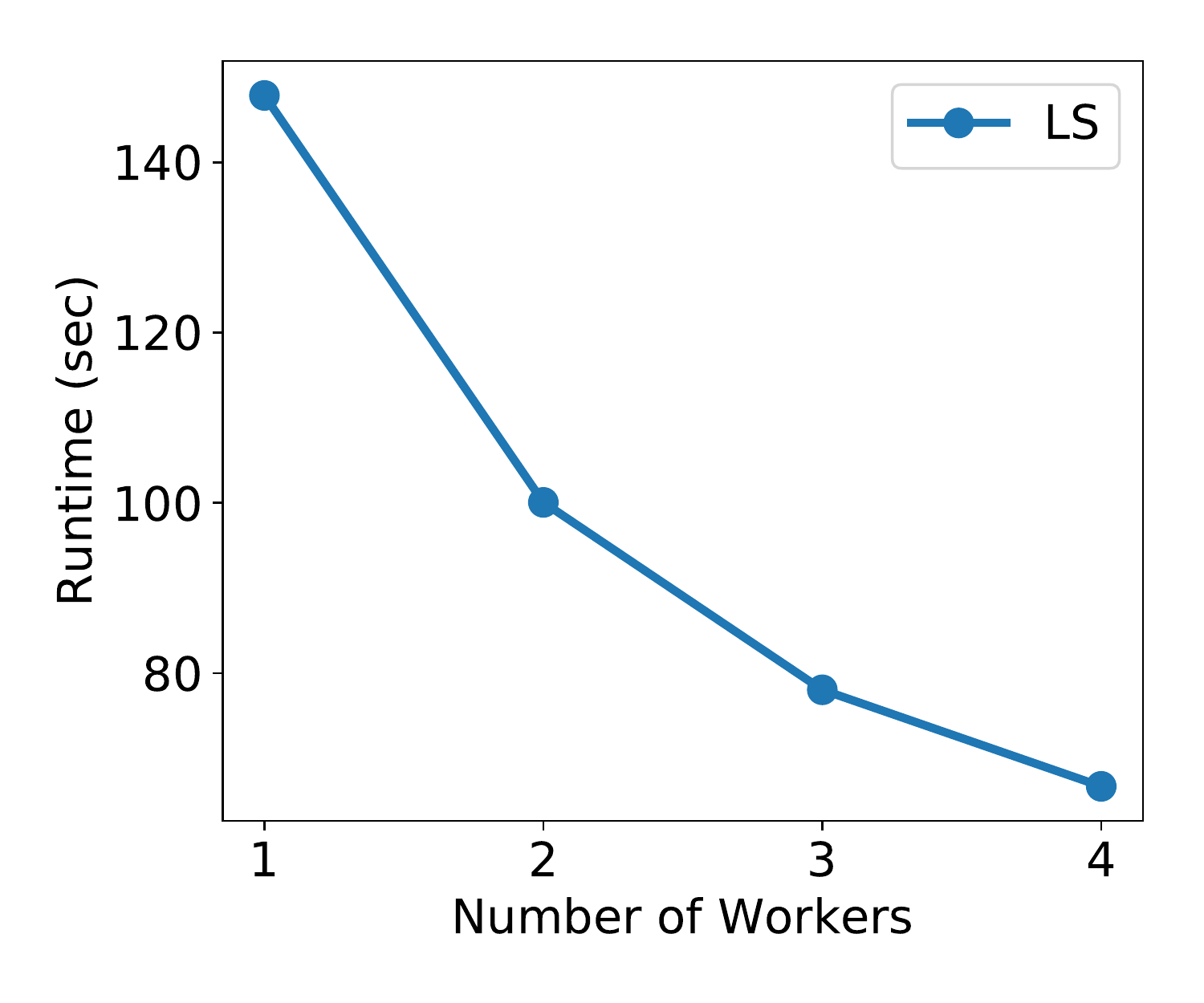}
     \caption{Parallelization}
\end{subfigure}
\begin{subfigure}{0.48\columnwidth}
    \includegraphics[width=\columnwidth]{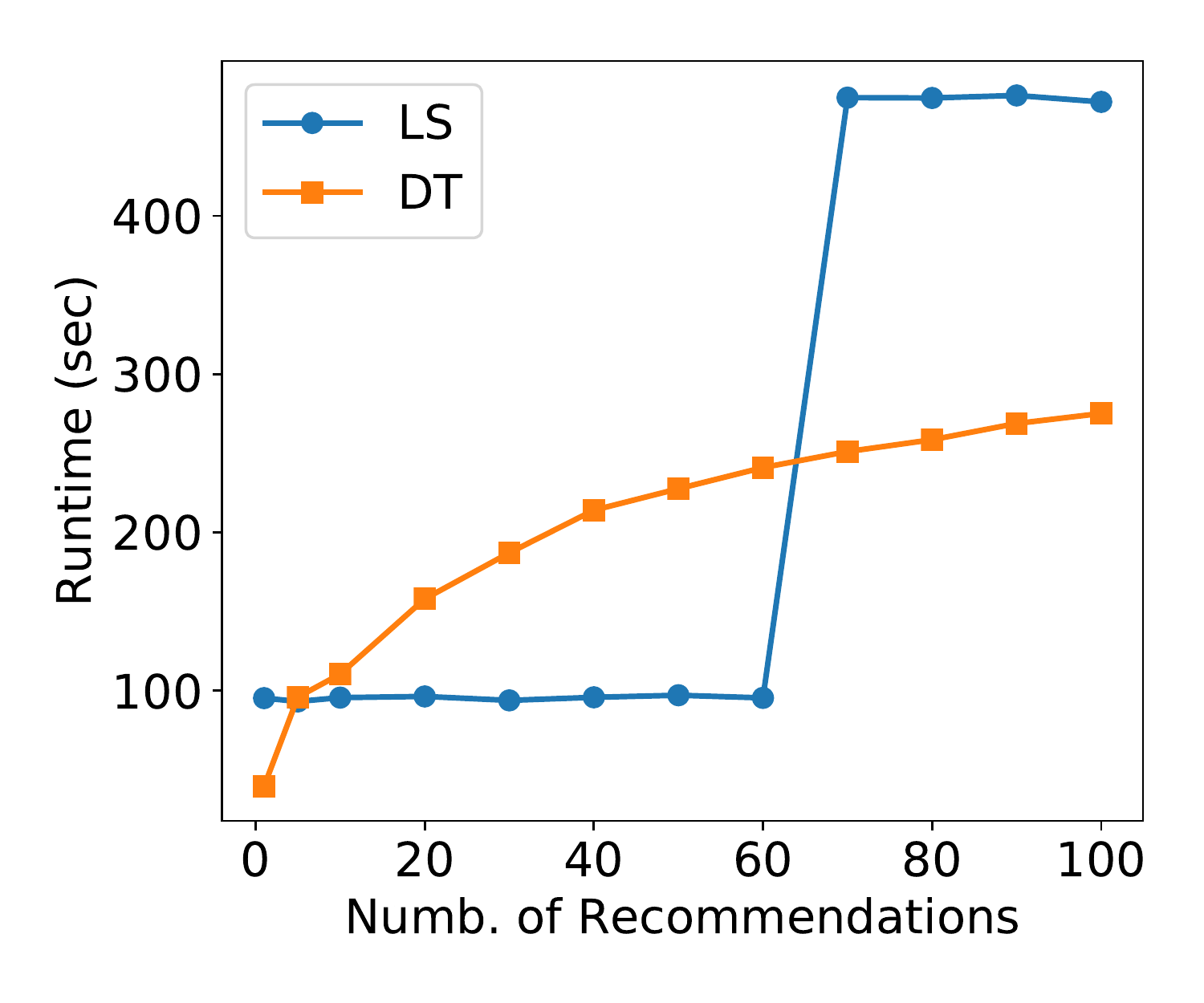}
    \caption{\# Recommendations}
\end{subfigure}
\caption{(a) \slicefinder{} runtime results with increasing number of (a) parallel workers and (b) recommendations (Census Income data).} 
\label{fig:parallel}
\end{figure}

Figure~\ref{fig:parallel}(a) illustrates how \slicefinder{} can scale with parallelization. LS can distribute the evaluation (e.g., effect size computation) of the slices with the same number of filter predicates to multiple workers. As a result, for the full Census Income data, increasing the number of workers results in better runtime. Notice that the marginal runtime improvement decreases as we add more workers. The  results for DT are not shown here because the current implementation does not support parallel DT model training.

Figure~\ref{fig:parallel}(b) compares the runtimes of LS and DT when the number of top-$k$ recommendations increase. For small $k$ values less than 5, DT is faster because it searchers fewer slices to find $k$ problematic ones. However, as $k$ increases, DT needs to search through many levels of a decision tree and starts to run relatively slower than LS. Meanwhile, LS only searches the next level of the lattice if $k$ is at least 70 at which point DT is again relatively faster. Thus, whether LS or DT is faster depends on $k$ and how frequently problematic slices occur.

\subsection{Interpretability}
\label{sec:interpretability}

\begin{table*}[t]
\centering
\begin{tabular}{| l | @{\hspace{2pt}}c@{\hspace{2pt}} |@{\hspace{2pt}} c @{\hspace{2pt}}|@{\hspace{2pt}} c @{\hspace{2pt}}|}
\hline
\hspace{7cm}{\bf Slice} & {\bf \# Literals} & {\bf Size} & {\bf Effect Size} \\
\hline \hline
\multicolumn{4}{|c|}{LS slices from Census Income data} \\
\hline
\predicate{Marital Status}{Married-civ-spouse} & 1 & 14065 & 0.58  \\
\predicate{Relationship}{Husband} & 1 & 12463 & 0.52  \\
\predicate{Relationship}{Wife} & 1 & 1406 & 0.46  \\
\predicate{Capital Gain}{3103} & 1 & 94 & 0.87  \\
\predicate{Capital Gain}{4386} & 1 & 67 & 0.94  \\
\hline 
\multicolumn{4}{|c|}{DT slices from Census Income data} \\
\hline
\predicate{Marital Status}{Married-civ-spouse} & 1 & 14065 & 0.58 \\
{\sf Marital Status $\neq$ Married-civ-spouse} $\rightarrow$ {\sf Capital Gain $\geq$ 7298} $\rightarrow$ {\sf Capital Gain $<$ 8614} $\rightarrow$ {\sf Education-Num $<$ 13} & 4 & 7 & 0.58\\
{\sf Marital Status $\neq$ Married-civ-spouse} $\rightarrow$ {\sf Capital Gain $<$ 7298} $\rightarrow$ {\sf Education-Num $\geq$ 13}  $\rightarrow$ {\sf Age $\geq$ 28} $\ldots$  & 5 & 855 & 0.43\\
\ \ \ $\rightarrow$ {\sf Hours per week $\geq$ 44} & & & \\
{\sf Marital Status $\neq$ Married-civ-spouse} $\rightarrow$ {\sf Capital Gain $<$ 7298} $\rightarrow$ {\sf Education-Num $\geq$ 13} $\rightarrow$ {\sf Age $<$ 28} $\ldots$ & 5 & 5 & 1.07\\
\ \ \ $\rightarrow$ {\sf Capital Loss $\geq$ 2231} & & & \\
{\sf Marital Status $\neq$ Married-civ-spouse} $\rightarrow$ {\sf Capital Gain $<$ 7298} $\rightarrow$ {\sf Education-Num $\geq$ 13} $\rightarrow$ {\sf Age $\geq$ 28} $\ldots$ & 6  & 101 & 0.47\\
\ \ \ $\rightarrow$ {\sf Hours per week $<$ 44} $\rightarrow$ {\sf Education-Num $\geq$ 15} & & & \\
\hline 
\multicolumn{4}{|c|}{LS slices from Credit Card Fraud data} \\
\hline
\predicate{V14}{-3.69 -- -1.00} & 2 & 98 & 0.45  \\
\predicate{V7}{0.94 -- 23.48} $\wedge$ \predicate{V10}{-2.16 -- -0.87} & 3 & 29 & 0.41  \\
\predicate{V1}{1.13 -- 1.74} $\wedge$ \predicate{V25}{0.48 -- 0.71} & 4 & 28 & 0.54  \\
\predicate{V7}{0.94 -- 23.48} $\wedge$ \predicate{Amount}{270.54 -- 4248.34} & 4 & 28 & 0.53  \\
\predicate{V10}{-2.16 -- -0.87} $\wedge$ \predicate{V17}{0.92 -- 6.74} & 5 & 27 & 0.44  \\
\hline 
\multicolumn{4}{|c|}{DT slices from Credit Card Fraud data} \\
\hline
{\sf V14} $< -2.17\rightarrow$ {\sf V10} $\geq -1.52$ & 2 & 31 & 0.60 \\
{\sf V14} $\geq -2.17\rightarrow$ {\sf V4} $\geq 0.76\rightarrow$ {\sf V12} $< -0.42$ & 3 & 59 & 0.48 \\
{\sf V14} $\geq -2.17\rightarrow$ {\sf V4} $< 0.76\rightarrow$ {\sf V14} $< -0.93\rightarrow$ {\sf V2} $< 1.04$ & 4  & 23 &0.42\\
{\sf V14} $\geq -2.17\rightarrow$ {\sf V4} $< 0.76\rightarrow$ {\sf V14} $\geq -0.93\rightarrow$ {\sf Amount} $\geq 320$ & 4 & 18 &0.52\\
{\sf V14} $\geq -2.17\rightarrow$ {\sf V4} $\geq 0.76\rightarrow$ {\sf V12} $\geq -0.42\rightarrow$ {\sf Amount} $\geq 1\rightarrow$ {\sf V17} $\geq 1.68$ & 5 & 6 &0.63\\
\hline
\end{tabular}
\caption{Top-5 slices found by LS and DT from the Census Income and Credit Card Fraud datasets. When denoting a slice from a decision tree, we use the $\rightarrow$ notation to order the literals by level.}
\label{tbl:app-top-slices}
\end{table*}

An important feature of \slicefinder{} is that it can find interpretable slices that can help a user understand and describe the model's behavior using a few common features. A user without \slicefinder{} may have to go through all the misclassified examples (or clusters of them) manually to see if the model is biased or failing.

Table~\ref{tbl:app-top-slices} shows 
top-5 problematic slices from the two datasets using LS and DT. Looking at the top-5 slices found by LS from the Census Income data, the slices are easy to interpret with a few number of common features. We see that the \predicate{Marital Status}{Married-civ-spouse} slice has the largest size as well as a large effect size, which indicates that the model can be improved for this slice. It is also interesting to see that the model fails for the people who are husbands or wives, but not for other relationships: own-child, not-in-family, other-relative, and unmarried. We also see slices with high capital gains tend to be problematic in comparison to the common case where the value is 0. In addition, the top-5 slices found by DT from the Census Income data can also be interpreted in a straightforward way, although having more literals makes the interpretation more tedious. Finally, the top-5 slices from the Credit Card Fraud data are harder (but still reasonable) to interpret because many feature names are anonymized (e.g., V14).

\subsection{False Discovery Control}
\label{sec:falsediscoverycontrol}

\begin{figure}[t]
\centering
\begin{subfigure}{0.48\columnwidth}
     \includegraphics[width=\columnwidth]{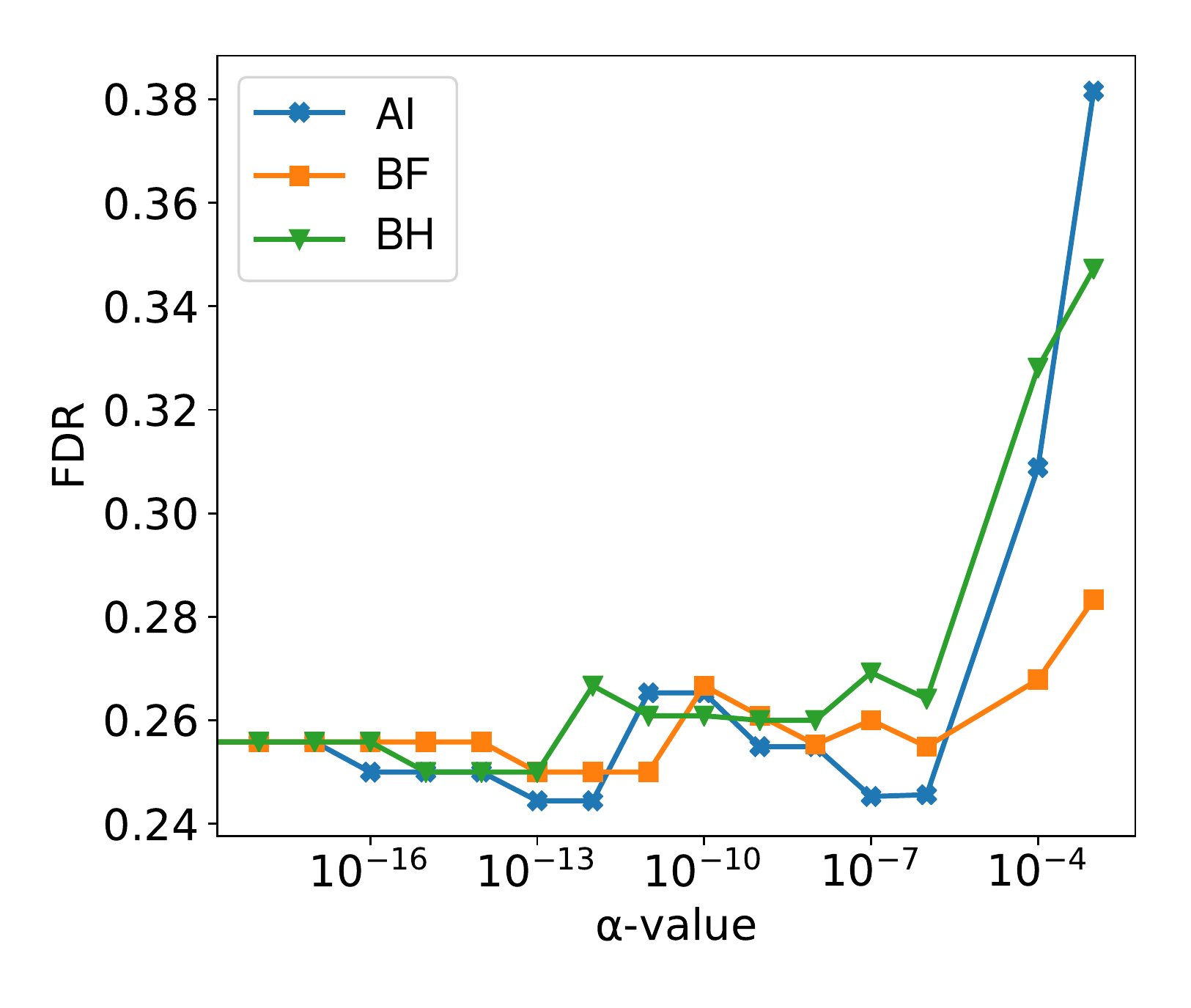}
     \caption{False Discovery Rate}
\end{subfigure}
\begin{subfigure}{0.48\columnwidth}
     \includegraphics[width=\columnwidth]{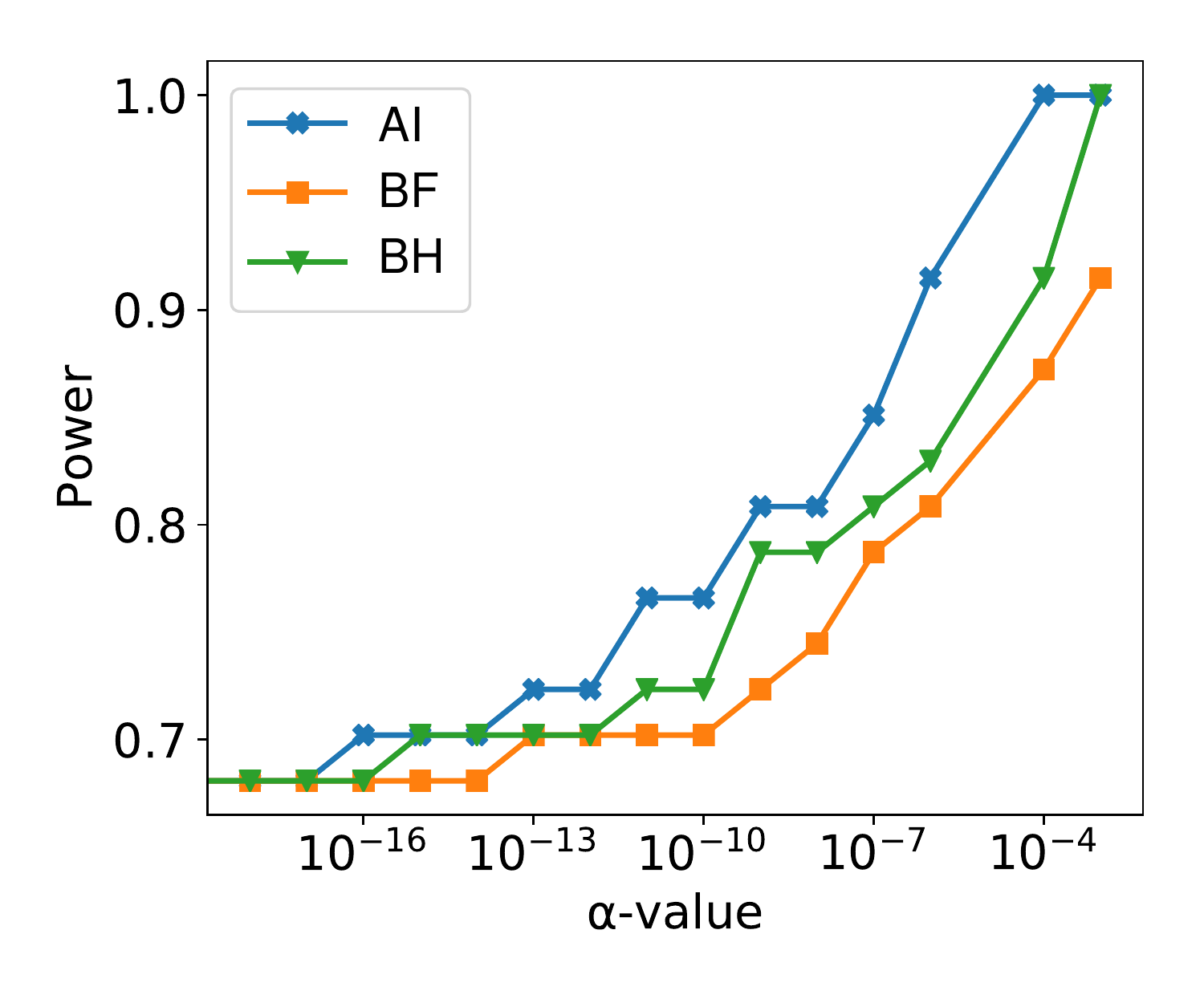}
     \caption{Power}
\end{subfigure}
\caption{(a) False discovery rate and (b) power comparison of the Bonferroni, Benjamini Hochberg, and $\alpha$-investing techniques (Census Income data).}     
\label{fig:falsediscovery}
\end{figure}

Even for a small data set (or sample), there can be an overwhelming number of problematic slices. The goal of \slicefinder{} is to bring the user's attention to a handful of large problematic slices; however, if the sample size is small, most slices would contain fewer examples, and thus, it is likely that many slices and their effect size measures are seen by chance. 
In such a case, it is important to prevent false discoveries (e.g., non-problematic slices appear as problematic where $\phi \geq T$ due to sampling bias). For evaluation, we use the Census Income data and compare the results of Bonferroni correction (BF), the Benjamini-Hochberg procedure (BH), and $\alpha$-investing (AI) using two standard measures: {\em false discovery rate}, which was described in Section~\ref{sec:falsediscoverycontrol}, and {\em power}~\cite{DBLP:conf/sigmod/ZhaoSZBUK17}, which is the probability that the tests correctly reject the null hypothesis.

The results in Figure~\ref{fig:falsediscovery} show that, as the $\alpha$-value (or wealth when using AI) increases up to 0.01, AI and BH have higher FDR results than BH, but higher power results as well. When measuring the accuracy of slices, AI slightly outperforms both BH and BF because it invests its $\alpha$ more effectively using the Best-foot-forward policy. In comparison, BF is conservative and has a high false-discovery rate (which results in lower accuracy), and BH does not exploit the fact that the earlier slices are more likely to be problematic as AI does. The more important advantage of AI is that it is the only technique that works in an interactive setting.

\section{Related Work}

In practice, the overall performance metrics can mask the issues on a more granular-level, and it is important to validate the model accordingly on smaller subsets/sub-populations of data (slices). While a well-known problem, the existing tools are still primitive in that they rely on domain experts to pre-define important slices.
State-of-art tools for machine learning model validation include Facets~\cite{facets}, which can be used to discover bias in the data, TensorFlow Model Analysis (TFMA), which slices data  by an input feature dimension for a more granular performance analysis~\cite{tfma}, and MLCube~\cite{kahng2016visual}, which provides manual exploration of slices and can both evaluate a single model or compare two models. While the above tools are manual, \slicefinder{} complements them by automatically finding slices useful for model validation.

%\slicefinder{} can be used along with visualization tools for validating the mode.
%Facets Overview~\cite{facets} is a general tool that displays feature-wise numeric and categorical statistics.
%MLCube~\cite{kahng2016visual} provides manual exploration of slices and can both evaluate a single model or compare two models. 
%t-SNE is a widely-used visualization tool for clustering embeddings. In comparison, \slicefinder{} aims to automatically find the interesting slices.

There are several other lines of work related to this problem, and 
 we list the most relevant work to \slicefinder{}.

\noindent\textbf{Data Exploration: }
Online Analytical Processing (OLAP) has been tackling the problem of slicing data for analysis, and the techniques deal with the problem of large search space (i.e., how to efficiently identify data slices with certain properties). For example, Smart Drilldown~\cite{joglekar2016interactive} proposes an OLAP drill down process that returns the top-$k$ most ``interesting'' rules such that the rules cover as many records as possible while being as specific as possible. Intelligent rollups~\cite{sathe2001intelligent} goes the other direction where the goal is to find the broadest cube that share the same characteristics of a problematic record. In comparison, \slicefinder{} finds slices, on which the model underperforms, without having to evaluate the model on all the possible slices. This is different from general OLAP operations based on cubes with pre-summarized aggregates, and the OLAP algorithms cannot be directly used.

\noindent\textbf{Model Understanding: } 
Understanding a model and its behavior is a broad topic that is being studied extensively~\cite{Freitas:2014:CCM:2594473.2594475,ribeiro2016explaining,anchors:aaai18,tamagnini2017interpreting,bastani2017interpreting,lakkaraju2017interpretable}. For example, LIME~\cite{ribeiro2016explaining} trains interpretable linear models on local data and random noise to see which feature are prominent. Anchors~\cite{anchors:aaai18} are high-precision rules that provide local and sufficient conditions for a black-box model to make predictions. In comparison, \slicefinder{} is a complementary tool to provide part of the data where the model is performing relatively worse than other parts. As a result, there are certain applications (e.g., model fairness) that benefit more from slices. PALM~\cite{krishnan2017palm} isolates a small set of training examples that have the greatest influence on the prediction by approximating a complex model into an interpretable meta-model that partitions the training data and a set of sub-models that approximate the patterns within each pattern. PALM expects as input the problematic example and a set of features that are explainable to the user. In comparison, \slicefinder{} finds large, significant, and interpretable slices without requiring user input.
Influence functions~\cite{koh2017understanding} have been used to compute how each example affects model behavior. In comparison, \slicefinder{} identifies interpretable slices instead of individual examples. An interesting research direction is to extend influence functions to slices and quantify the impact of slices on the overall model quality.

\noindent\textbf{Feature Selection: }
\slicefinder{} is a model validation tool, which comes after model training. It is important to note that this is different from feature selection \cite{charikar2000combinatorial,guyon2003introduction} in model training, where the goal is often to identify and (re-)train on the most correlated features (dimensions) to the target label (i.e., finding representative features that best explain model predictions). Instead, \slicefinder{} identifies a few common feature values that describe subsets of data with significantly high error concentration for a given model; this, in turn, could help the user to interpret hidden model performance issues that are masked by good overall model performance metrics.

\section{Conclusion}

We have proposed \slicefinder{} as a tool for efficiently and accurately finding large, significant, and interpretable slices. The techniques are relevant to model validation in general, but also to model fairness and fraud detection where human interpretability is critical to understand model behavior. We have proposed two complementing approaches for slice finding: decision tree training, which finds non-overlapping slices, and lattice searching, which finds possibly-overlapping slices. We also provide an interactive visualization front-end to help users quickly browse through a handful of problematic slices.

In the future, we would like to improve \slicefinder{} to better discretize numeric features and support the merging and summarization of slices. We would also like to deploy \slicefinder{} to production machine learning platforms and conduct a user study on how helpful the slices are for explaining and debugging models.

\section*{Acknowledgments}

Steven Euijong Whang and Ki Hyun Tae were supported by a Google AI Focused Research Award and by the Engineering Research Center Program through the National Research Foundation of Korea (NRF) funded by the Korean Government MSIT (NRF-2018R1A5A1059921).

\bibliographystyle{IEEEtran}
\bibliography{main}

\end{document}